\documentclass[aps,prd,twocolumn,showpacs,groupedaddress,nofootinbib]{revtex4}

\usepackage{graphicx}  % needed for figures
\usepackage{dcolumn}   % needed for some tables
\usepackage{bm}        % for math
\usepackage[english]{babel}
\usepackage{amsfonts,amsmath,amssymb,mathrsfs}
\usepackage{times}
\newcommand{\eq}{\begin{equation}}
\newcommand{\eeq}{\end{equation}}
\newcommand\T{\rule{0pt}{3ex}}
\newcommand\B{\rule[-2ex]{0pt}{0pt}}
\newcommand{\mnras}{Mon.\ Not.\ Roy.\ Astron.\ Soc.\ }
\newcommand{\apjl}{Astrophys.\ J.\ Lett.\ }
\newcommand{\aap}{Astron.\ Astrophys.\ }
\newcommand{\cqg}{Class.\ Quant. Grav.\ }

\begin{document}

\title{The influence of the hydrodynamic drag from an accretion torus on
extreme mass-ratio inspirals}

\author{Enrico Barausse}\email{barausse@sissa.it}
\affiliation{SISSA, International School for
             Advanced Studies and INFN, Via Beirut 2, 34014 Trieste, Italy}

\author{Luciano Rezzolla}%\email{rezzolla@aei.mpg.de}
\affiliation{Max-Planck-Institut f\"ur Gravitationsphysik,
             Albert-Einstein-Institut, 14476 Potsdam, Germany}

\affiliation{INFN, Sezione di Trieste, Via A. Valerio 2, I-34127 Trieste, Italy}

\date{\today}
\begin{abstract}
	We have studied extreme mass-ratio inspirals (EMRIs) in
	spacetimes containing a rotating black hole and a non
	self-gravitating torus with a constant distribution of
	specific angular momentum. We have found that the dissipative
	effect of the hydrodynamic drag exerted by the torus on the
	satellite is much smaller than the corresponding one due to
	radiation reaction, for systems such as those generically
	expected in Active Galactic Nuclei and at distances from the
	central supermassive black hole (SMBH) which can be probed
	with LISA. However, given the uncertainty on the parameters of
	these systems, namely on the masses of the SMBH
	and of the torus, as well as on its size, there exist
	configurations in which the effect of the hydrodynamic drag on
	the orbital evolution can be comparable to the
	radiation-reaction one in phases of the inspiral which are
	detectable by LISA.  This is the case, for instance, 
        for a $10^6M_{\odot}$ SMBH surrounded by a corotating torus of 
        comparable mass and with radius of $10^3-10^4$ gravitational radii,
        or for a $10^5 M_{\odot}$ SMBH surrounded by a corotating $10^4 M_{\odot}$ torus
	with radius of $10^5$ gravitational radii. Should these
	conditions be met in astrophysical systems, EMRI-gravitational
	waves could provide a characteristic signature of the
	presence of the torus. In fact, while radiation reaction
	always increases the inclination of the orbit with respect to
	the equatorial plane (\textit{i.e.,} orbits evolve towards the
	equatorial retrograde configuration), the hydrodynamic drag
	from a torus corotating with the SMBH always decreases
	it (\textit{i.e.,} orbits evolve towards the equatorial
	prograde configuration). However, even when initially
	dominating over radiation reaction, the influence of the
	hydrodynamic drag decays very rapidly as the satellite moves
	into the very strong-field region of the SMBH (\textit{i.e.,}
	$p\lesssim5M$), thus allowing one to use pure-Kerr templates
	for the last part of the inspiral. Although our results have
 	been obtained for a specific class of tori, we argue that they
 	will be qualitatively valid also for more generic distributions of the specific
 	angular momentum.
\end{abstract}

\pacs{04.30.-w,04.70.-s,98.35.Jk,98.62.Js}
\maketitle

\section{\label{sec:intro}Introduction}

One of the most exciting prospects opened up by the scheduled launch
of the space-based gravitational-wave detector LISA~\cite{LISA} will
be the possibility of mapping accurately the spacetime of the
supermassive black holes (SMBHs) which are believed to reside in the
center of galaxies~\cite{SMBH}. Among the best candidate sources for
this detector there are Extreme Mass Ratio Inspirals (EMRIs),
\textit{i.e.} stellar-mass black holes ($m\approx1-10M_\odot$) or
compact objects orbiting around the SMBH and slowly
inspiraling due to the loss of energy and angular momentum via
gravitational waves (radiation reaction).  In order for the signal to
fall within the sensitivity band of LISA, the SMBH must have a mass
$M\approx10^5-10^7M_\odot$, \textit{i.e.}, the low end of the SMBH
mass function.

It is currently expected that a number of such events ranging from
tens to perhaps one thousand could be measured every year
\cite{gair_event_rates}, but since they will have small
signal-to-noise ratios, their detection and subsequent parameter
extraction will require the use of matched-filtering techniques.  These
basically consist of cross-correlating the incoming gravitational-wave
signal with a bank of theoretical templates representing the expected
signal as a function of the parameters of the source.

This will not only allow one to detect the source, but also to extract
its properties. For instance, the accurate modelling of the motion of
a satellite in a Kerr spacetime will allow one to measure the spin and
the mass of the SMBH.  Although producing these pure-Kerr
templates has proved to be a formidable task, particularly because of
the difficulty of treating rigorously the effect of radiation reaction
(see Ref.~\cite{Poisson_review} for a detailed review), considerable
effort has gone into trying to include the effects of a deviation from
the Kerr geometry. These attempts are motivated by the fact that
possible ``exotic'' alternatives to SMBHs have been proposed
(\textit{e.g.,} boson stars~\cite{monica}, fermion balls
\cite{fermion_balls} and gravastars~\cite{gravastars}), although the
presence of these objects would require to modify radically the
mechanism with which galaxies are expected to form. On the other hand,
non-pure Kerr templates might allow one to really map the spacetimes
of SMBHs and to test, \textit{experimentally}, the Kerr solution.

Different approaches to this problem have been considered in the
literature. EMRIs in a spacetime having arbitrary gravitational
multipoles should be considered in order to maintain full generality
\cite{ryan}, but this method does not work very well in practice and
would only apply to vacuum spacetimes.  For this reason, alternative
approaches have been proposed and range from EMRIs around non-rotating
boson stars~\cite{gair_boson}, to EMRIs in bumpy black-hole spacetimes
\cite{bumpyBH} (\textit{i.e.}, spacetimes which are \textit{almost}
Schwarzschild and require naked singularities or exotic matter) or in
quasi-Kerr spacetimes~\cite{quasi_kerr} (\textit{i.e.}, spacetimes,
consisting of Kerr plus a small quadrupole moment).

Interestingly, none of these methods is suitable for taking into
account the effect of the matter which is certainly present in
galactic centers. SMBHs can indeed be surrounded by stellar disks (as
in the case of the Galactic center~\cite{stellar_disks}) or, as in the
case of Active Galactic Nuclei (AGNs)~\cite{agn}, in which we are most interested, by
accretion disks of gas and dust which can be even as massive as the
SMBH~\cite{Hure}. While the gravitational attraction of a disk
can have important effects on EMRIs if this disk is very massive and
close to the SMBH~\cite{BHTorus}, an astrophysically realistic
accretion disk can influence an EMRI only if the satellite crosses it,
thus experiencing a ``hydrodynamic'' drag force.

This drag consists of two parts. The first one is due to the accretion
of matter onto the satellite black hole (this was studied analytically
by Bondi \& Hoyle~\cite{bondi} and subsequently confirmed through
numerical calculations~\cite{petrich,font1,font2}). This transfers energy and
momentum from the disk to the satellite, giving rise to a short-range
interaction.  The second one is instead due the gravitational
deflection of the material which is not accreted, which is
therefore far from the satellite, but which can nevertheless transfer
momentum to it. This long-range interaction can also be
thought of as arising from the gravitational pull of the satellite by its
own gravitationally-induced wake (\textit{i.e.,} the density
perturbations that the satellite excites, by gravitational
interaction, in the medium), and is often referred to as ``dynamical
friction''. This effect was first studied in a collisionless medium by
Chandrasekar~\cite{chandra}, but acts also for a satellite moving in a
collisional fluid
\cite{rephaeli,petrich,ostriker,sanchez,kim,my_dyn_frict}.

The effect of this disk-satellite interaction on EMRIs has been
studied by different authors for a number of disk models. 
More specifically, in a series of papers Karas, Subr and
Vokrouhlicky considered the interaction between stellar satellites and
thin disks \cite{karas_subr0,karas_subr1,karas_subr2}. In Ref.~\cite{karas_subr2},
in particular, Subr and Karas found that the effect of the star-disk
interaction on EMRIs dominates over radiation reaction for thin disks,
both for non-equatorial orbits crossing the disk only twice per
revolution and for equatorial orbits embedded in the disk. The only
exceptions to these conclusions come if the satellite is very compact
(a neutron star or a black hole) or the disk has a low density
(\textit{e.g.,} in the region close to the SMBH if the flow
becomes advection-dominated). These results agree with those found by
Narayan~\cite{narayan}, who focused on Advection Dominated Accretions
Flows (ADAFs), which were believed to describe accretion onto ``normal'' galactic
nuclei (\textit{i.e.,} ones much dimmer than AGNs)\footnote{Accretion onto ``normal''  galactic nuclei is now believed to be
better described by Advection Dominated Inflow Outflow Solutions
(ADIOS)~\cite{ADIOS}. However, this is not expected to change
significantly Narayan's results since ADIOS's, like ADAFs, have very
low densities in the vicinity of the SMBH.}. Overall, he found
that for compact objects and white dwarfs the effect of the
hydrodynamic drag is negligible with respect to radiation reaction,
whereas it is not negligible for main sequence and giant
stars. More recently, Levin~\cite{levin} has proposed a scenario in which massive stars form in a
thin accretion disk in an AGN, ultimately producing stellar-mass black
holes embedded in the disk.
The small black holes are then dragged towards the (non-rotating) SMBH, but if this is accreting at a rate comparable to the
Eddington limit, the drag from the accreting gas will not affect the final part
of the inspiral (\textit{i.e.,} at radii smaller than $10$ Schwarzschild radii) significantly.
Finally, Chakrabarti~\cite{chakra}, studied instead the orbital evolution
of a satellite black hole on a circular equatorial orbit embedded in a
disk with a non-Keplerian distribution of angular momentum, and found
that the exchange of angular momentum between the disk and the
satellite can lead to significant orbital modifications.

All of these studies have been been carried out within a Newtonian or
pseudo-Newtonian description of gravity (with the partial exception of Ref.~\cite{karas_subr0},
in which the orbits are Kerr geodesics, but
the disk model and the hydrodynamic drag is not relativistic). In this paper, instead, we
provide a first relativistic treatment for satellite  black holes
moving on generic orbits around a rotating SMBH surrounded by a thick
disk (\textit{i.e.,} a torus). We consider the torus to have constant
specific angular momentum and neglect its self-gravity (\textit{i.e.,}
we consider the metric to be pure Kerr). Under these assumptions, an
analytical solution exists for this
system~\cite{constant_l_disks1,constant_l_disks2}. This configuration
can be proved to be marginally stable with respect to axisymmetric
perturbations~\cite{marginally_stable} (\textit{i.e.,} if perturbed,
such a torus can accrete onto the SMBH), and is expected to be
a good approximation at least for the inner parts of the accretion
flow~\cite{constant_l_disks1,constant_l_disks2}.

We have found that for a system composed of a SMBH with mass
$M=10^6 M_\odot$ and a torus with mass $M_{\rm t}\lesssim M$ and
outer radius $r_{\rm out}=10^5 M$, the effect of the hydrodynamic drag
on the motion of the satellite black hole is much smaller than
radiation reaction at those distances from the SMBH which can be
probed with LISA (\textit{i.e.,} $\sim 10 M$ for $M=10^6
M_\odot$). Although these values for $M$, $M_{\rm t}$ and $r_{\rm
out}$ are plausible for AGNs, an overall uncertainty is still present
and has motivated an investigation also for different masses and sizes
of the torus. In this way we have found that the effect of the torus
can be important in the early part of the inspiral and that it could
leave an observable imprint in the gravitational waveforms detected by
LISA, if the radius of the torus is decreased to $r_{\rm
out}=10^3-10^4 M$ or, even for $r_{\rm out}=10^5 M$ and $M_{\rm
t}\lesssim M$, if $M=10^5 M_\odot$. In this latter case, in fact, LISA
could detect an EMRI event at distances as large as $r\sim 45 M$ from
the SMBH, although the event needs to be sufficiently close to
us because the amplitude of the gravitational-wave signal decreases as
$M/r$.

In addition, if non-negligible, the effect of the hydrodynamic drag
would have a distinctive signature on the waveforms. Radiation
reaction, in fact, always increases the inclination of the orbit with
respect to the equatorial plane (\textit{i.e.,} orbits evolve towards
the equatorial retrograde configuration)~\cite{NHSO}. The hydrodynamic drag from a
torus corotating with the SMBH, on the other hand, always
decreases this angle (\textit{i.e.,} orbits evolve towards the
equatorial prograde configuration). Should such a behavior be observed
in the data, it would provide a strong qualitative signature of the
presence of the torus. However, it is important to point out that even
for those configurations in which the hydrodynamic drag plays a major
role, this is restricted to the initial part of the inspiral
detectable by LISA, whereas its effect rapidly vanishes in the very
strong-field region of the SMBH (\textit{i.e.,} $p\lesssim5M$). As a
result, the pure-Kerr templates would provide a faithful description
of the last part of the inspiral even in these cases.

The rest of the paper is organized as follows. In Sec. \ref{sec:torus}
we review the equilibrium solutions that we used for the orbiting
torus. In Sec. \ref{sec:interaction} we present the equations
governing the interaction between the satellite black hole and the
torus, while in Sect.~\ref{sec:adiabatic} we apply the adiabatic
approximation to the hydrodynamic drag. Results are then discussed in
Sec. \ref{sec:eq_circ} for equatorial circular orbits and in
Sec. \ref{sec:generic} for generic (inclined and eccentric) orbits. 
Finally, the conclusions are drawn in
Sec.~\ref{sec:conclusion}. Throughout this paper we use units in which
$G=c=1$.

\section{\label{sec:torus}Modelling the torus}

The properties of non self-gravitating, stationary, axisymmetric and
plane-symmetric toroidal fluid configurations in Kerr spacetimes are
well-known in astrophysics but are less well known within the
community working on EMRIs. Because of this, in this Section we
briefly review the basic facts, referring the interested reader to
Refs.~\cite{constant_l_disks1,constant_l_disks2,font_daigne,zanotti_etal:03,zanotti_etal:05}
for additional information.

Let us consider a perfect fluid with 4-velocity $\boldsymbol{u}^{\rm
fluid}$, which is described by the stress-energy tensor
\begin{align}
\label{stress-tensor}
T^{\mu\nu}&= (\rho+p)u_{\rm fluid}^\mu u_{\rm fluid}^\nu+p g^{\mu\nu}\nonumber
        \\&= \rho_0 h u_{\rm fluid}^\mu u_{\rm fluid}^\nu+p g^{\mu\nu} \, ,
\end{align}
where $p$, $\rho_0$, $\rho$ and $h\equiv (p+\rho)/\rho_0$ are the
pressure, rest-mass density, energy density and specific enthalpy of
the fluid. In what follows we will model the fluid with a polytropic
equation of state $p=\kappa\rho_0^\Gamma=\rho_0\varepsilon(\Gamma-1)$,
where $\varepsilon=\rho/\rho_0-1$ is the internal energy per unit
rest-mass, and $\kappa$ and $\Gamma$ are the polytropic constant and
index, respectively. Because we are neglecting the self-gravity of the
fluid, we can also consider $\boldsymbol{g}$ as given by the Kerr
metric in Boyer-Lindquist coordinates, which reads~\cite{MTW}
\begin{eqnarray}
\label{Kerr}
ds^2 &=&
- \left( 1-\frac{2Mr}{\Sigma} \right) ~dt^2
+ \frac{\Sigma}{\Delta}~dr^2
+ \Sigma~d\theta^2 \nonumber \\
&&+ \left( r^2+a^2 + \frac{2Ma^2r}{\Sigma}\sin^2\theta \right)\sin^2\theta~d\phi^2 \nonumber \\
&&- \frac{4Mar}{\Sigma}\sin^2\theta~dt~d\phi\,,
\end{eqnarray}
where
\begin{align}
& \Sigma \equiv r^2 + a^2\cos^2\theta, & \Delta \equiv r^2 - 2Mr + a^2. &
\end{align}

The fluid is assumed to be in circular non-geodesic motion with 4-velocity
\begin{align}
&\boldsymbol{u}^{\rm fluid}=A(r,\theta)\left[\frac{\partial}{\partial t}+
\Omega(r,\theta)\frac{\partial}{\partial \phi}\right]\nonumber\\
&=U(r,\theta)\left[-dt+\ell(r,\theta) d\phi\right]\;,\label{eq:u}
\end{align}
where the second equals sign underlines that the vector and the 1-form
are each the dual of the other. Here, $\Omega\equiv u_{\rm fluid}^\phi/u_{\rm fluid}^t$ is the angular
velocity, $A\equiv u_{\rm fluid}^t$ is called the redshift factor,
$U\equiv -u^{\rm fluid}_t$ is the energy per unit mass as measured at
infinity and $\ell \equiv -u^{\rm fluid}_\phi/u^{\rm fluid}_t$ is the
specific angular momentum as measured at infinity (\textit{i.e.,} the
angular momentum per unit energy as measured at infinity). Note that $\ell$ is
conserved for stationary axisymmetric flows, as can be easily shown
using Euler's equation. The specific angular momentum and the angular
velocity are trivially related by
\begin{equation}
\Omega=-\frac{g_{t\phi}+g_{tt}\ell}{g_{\phi\phi}+g_{t\phi}\ell}\,,\quad
\ell=-\frac{g_{t\phi}+g_{\phi\phi}\Omega}{g_{tt}+g_{t\phi}\Omega}\;,
\end{equation}
while the normalization condition $\boldsymbol{u}^{\rm fluid}\cdot\boldsymbol{u}^{\rm fluid}=-1$ gives
\begin{gather}
U=\sqrt{\frac{\varpi^2}{g_{tt}\ell^2+2 g_{t\phi}\ell+g_{\phi\phi}}}\label{eq:U}\;,\\
A=\sqrt{\frac{-1}{g_{tt}+2 g_{t\phi}\Omega+g_{\phi\phi}\Omega^2}}\;,\label{eq:A}\\
AU=\frac{1}{1-\Omega\ell}\label{eq:AU}\;,
\end{gather}
where $\varpi^2=
g_{t\phi}^2-g_{tt}\,g_{\phi\phi}=\Delta \sin^2\theta$. Note that in
this paper we will always consider $\ell>0$ (torus rotating in the
positive $\phi$-direction), while we will allow the spin parameter
$a$ of the black hole to be either positive (black hole corotating
with the torus) or negative (black hole counter-rotating with respect
to the torus).

To calculate the structure of the torus, we need to use Euler's
equation, which in its general form reads \eq\label{eq:euler} a_{\rm
fluid}^\mu=-\frac{(g^{\mu\nu}+u_{\rm fluid}^\mu u_{\rm
fluid}^\nu)\partial_\nu p}{p+\rho}\;, \eeq where $a_{\rm fluid}^\mu$
is the 4-acceleration of the fluid. In particular, if the pressure is
assumed to depend only on $r$ and $\theta$ and if the equation of
state is barotropic [\textit{i.e.,} if $\rho=\rho(p)$]\footnote{This
is of course the case for a polytropic equation of state, because
$\rho=p/(\Gamma-1)+(p/\kappa)^{1/\Gamma}$.}, from Eq. \eqref{eq:euler}
one easily gets that the 4-acceleration can be expressed as the
gradient of a scalar potential $W(p)$:
\eq\label{eq:W_def}
a^{\rm fluid}_\mu=\partial_\mu W \;, \quad W(p)= -\int^p
\frac{dp'}{p'+\rho(p')}\;. \eeq 
On the other hand, from the definition
of 4-acceleration ($a_{\rm fluid}^\mu=u_{\rm fluid}^\nu\nabla_\nu u_{\rm fluid}^\mu$),
Eqs. \eqref{eq:u}, \eqref{eq:A} and \eqref{eq:AU}, and the Killing
equation $\nabla_{(\mu}\xi_{\nu)}=0$ for $\xi=\partial /\partial t$
and $\xi=\partial /\partial \phi$, one easily gets
\eq\label{eq:euler2} a^{\rm fluid}_\mu=\partial_\mu W= -\frac{\partial_\mu
p}{p+\rho}= \partial_\mu \ln
U-\frac{\Omega}{1-\Omega\ell}\partial_\mu\ell\;. 
\eeq

In particular, taking the derivative of this equation,
anti-symmetrizing and using the trivial fact that
$\partial_{[\mu\nu]}W=\partial_{[\mu\nu]}\ell=\partial_{[\mu\nu]}U=0$,
we obtain that $\partial_{[\mu}\Omega\,\partial_{\nu]}\ell=0$. This
implies $\nabla\Omega\propto\nabla\ell$ and thus that $\ell$ and
$\Omega$ have the same contour levels [\textit{i.e.,}
$\Omega=\Omega(\ell)$]. Using this fact, we can then write
Eq. \eqref{eq:euler2} in an integral form:
\begin{align}
&W-W_{\rm out}= -\int_0^p \frac{dp'}{p'+\rho(p')}\nonumber\\& =\ln U
-\ln U_{\rm out} -\int_{\ell_{\rm
out}}^{\ell}\frac{\Omega(\ell')d\ell'}{1-\Omega(\ell')\ell'}\;,
\label{eq:euler_int}
\end{align}
where $W_{\rm out}$ and $\ell_{\rm out}$ are the potential and
specific angular momentum at the outer edge of the
torus.\footnote{Of course, $W_{\rm out}$ and $\ell_{\rm out}$ can be replaced
by the values of $W$ and $\ell$ at the inner edge of the torus if
this is present.}

In the case of a torus with constant specific angular momentum
[\textit{i.e.,} $\ell(r,\theta)=$ constant], Eq. \eqref{eq:euler_int}
 provides an analytical solution, because   
once $\ell$ has been fixed the integral on the right-hand
side is zero and Eq. \eqref{eq:U} gives an analytical expression
for $U$:
\begin{equation}
W-W_{\rm out}= -\int_0^p \frac{dp'}{p'+\rho(p')}=
\ln U -\ln U_{\rm out}\;.\label{eq:euler_int2}
\end{equation}
Note that if one requires that $W\to0$ when $r\to+\infty$ (\textit{i.e.,} $W=0$ for an equipotential surface
closing at infinity), this equation gives $W=\ln U$: $W>0$ then
corresponds to open equipotential surfaces, while $W<0$ corresponds to
closed equipotential surfaces. Interestingly, the potential well can
present a minimum and a saddle point. Because of the plane-symmetry,
these points are located in the equatorial plane, thus corresponding
to local extremes of $W(r,\theta=\pi/2)$, and mark two important
positions: respectively, the center of the torus (\textit{i.e.,} the
point where the density reaches its maximum) and its cusp
(\textit{i.e.,} the mass-shedding point). Noticeably, these points are
located at the radii where the specific angular momentum of the torus,
$\ell$, coincides with that of the geodesic circular equatorial orbit
(the ``Keplerian'' orbit) corotating with the torus,
\eq\label{eq:ell_Kepler} \ell_K(r,a)=\frac{r^{2} - 2 a
\sqrt{Mr} + a^{2}}{(r-2M)\sqrt{r/M} + a}\;. \eeq
This immediately follows from the fact that at the extremes of the function $W$
one has $\partial_\mu W=0$, which leads, through
Eq. \eqref{eq:W_def}, to $a_{\rm fluid}^\mu=0$ (in other words, at the
center and at the cusp the pressure gradients are zero and only
gravitational forces act).

In this paper we will indeed consider constant-$\ell$ tori. A detailed
classification of these models depending upon the values of $\ell$ and
of $W_{\rm out}$ can be found in
Refs.~\cite{constant_l_disks1,constant_l_disks2,font_daigne}.  Here we
simply recall that in order to have a closed equipotential surface
with a cusp, one needs to have a value of $\ell$ between the specific
angular momenta $\ell_{\rm ms}$ and $\ell_{\rm mb}$ of the marginally
stable and marginally bound equatorial geodesic (\textit{i.e.,}
``Keplerian'') orbits corotating with the torus. This can be easily
understood by noting, from Eq.~\eqref{eq:euler_int2}, 
that the potential $W(r,\theta=\pi/2)$ is simply the effective potential 
describing the equatorial motion of a test particle around a Kerr black hole.
As such,  $\ell_{\rm ms}$ and $\ell_{\rm mb}$ can be calculated easily using
Eq. \eqref{eq:ell_Kepler} and the formulas for the radii of the
marginally stable and marginally bound circular equatorial orbits in
Kerr rotating in the positive $\phi$-direction (\textit{i.e.,}
corotating with the torus):
\begin{align}
& \ell_{\rm ms}=\ell_K(r_{\rm ms})\,,\quad \ell_{\rm mb}=\ell_K(r_{\rm mb})\,,\label{eq:l_ms_l_mb}\\
& r_{\rm ms}/M=3+Z_{2}-{\rm sign}(\tilde{a})\sqrt{(3-Z_{1})(3+Z_{1}+2 Z_{2})}\,,\\ 
&r_{\rm mb}/M=2-\tilde{a}+2\sqrt{1-\tilde{a}}\,,\\
& Z_{1}=1+(1-\tilde{a}^{2})^{1/3} \left[ (1+\tilde{a})^{1/3}+(1-\tilde{a})^{1/3} \right]\,,\\
&Z_{2}=\sqrt{ 3 \tilde{a}^{2} + Z_{1}^{2}}\;,
\end{align}
where $\tilde{a}= a/M$.

In order to pick up a particular solution having both an inner and an
outer radius, one needs also to choose a negative value for the
``potential barrier'' at the inner edge of the torus, \eq \Delta
W=W_{\rm in}- W_{\rm cusp}=W_{\rm out}- W_{\rm cusp}\leq0\,.\eeq If
$\Delta W<0$, the inner radius of the torus is larger than the radius
at which the cusp occurs ($r_{\rm in}>r_{\rm cusp}$), while if the
potential barrier $\Delta W$ reduces to zero, the torus exactly fills
its outermost closed equipotential surface and $r_{\rm in}=r_{\rm
cusp}\leq r_{\rm ms}$. Note that because of the considerations that we
have made above about the value of $\ell$, for constant-$\ell$ tori we
have $r_{\rm cusp}\geq r_{\rm mb}$ (with $r_{\rm cusp}= r_{\rm mb}$
only if $\ell=\ell_{\rm mb}$) and $r_{\rm center}\geq r_{\rm ms}$
(with $r_{\rm center}= r_{\rm ms}$ only if $\ell=\ell_{\rm ms}$). If
instead $\Delta W>0$, the fluid overflows the outermost closed
equipotential surface and mass transfer is possible at the cusp: for a
polytropic equation of state, the accretion rate can be shown to be
$\dot M\propto \Delta W^{\Gamma/(\Gamma-1)}$.

The integral Euler equation for constant-$\ell$ tori [Eq. \eqref{eq:euler_int2}]
further simplifies if the equation of state is
polytropic, because in this case 
\eq\label{eq:euler_int3} \int_0^p
\frac{dp'}{p'+\rho(p')}=\ln\frac{h}{h_{\rm out}} \,,\eeq
where $h_{\rm out}$
is the specific enthalpy at the outer edge of the torus. Since for
a polytropic equation of state the enthalpy is given by
\begin{equation}
h = 1+\frac{\Gamma}{\Gamma-1}\kappa \rho_0^{\Gamma-1}\,,
\label{eq:h}
\end{equation}
it is clear that $h_{\rm out}=1$ (because $p=\rho_0=0$ at the outer
edge of the torus), and Eqs. \eqref{eq:euler_int2}
and \eqref{eq:euler_int3} give 
\eq \rho_0(r,\theta) =
\left\{\frac{\Gamma-1}{\Gamma}\frac{\left[e^{W_{\rm
out}-W(r,\theta)}-1\right]} {\kappa}\right\}^{1/(\Gamma-1)}\;.
\label{eq:rho} \eeq 
Once the rest-mass distribution is known, the
total rest mass of the torus is given by
\begin{equation}
M_{t,0} = \int\rho_0\sqrt{-g}u^t d^3x \ ,
\end{equation}
where $\sqrt{-g}=\Sigma\sin\theta$ and $d^3 x=dr \, d\theta\, d\phi$
is the coordinate 3-volume element, while the mass-energy reads
\begin{align}
&M_{\rm t} = \int (T^r_r + T^\phi_\phi +
        T^\theta_\theta - T^t_t)\sqrt{-g}\,d^3x=\nonumber\\
& 2\pi\int_{\rho_0 > 0}\left(\frac{g_{\phi\phi}-g_{tt} \ell^{2}}
{g_{\phi\phi}+2 g_{t\phi} \ell +g_{tt} \ell^{2}}\rho_0 h + 2 P\right)\nonumber\\
& \times\left(r^{2}+a^{2}\cos^{2}{\theta}\right)\sin{\theta}\ dr d\theta\;.
\end{align}
Clearly, the smaller the ratio between the mass of the torus and that
of the SMBH, the better the approximation of
neglecting the self-gravity of the torus.

\section{Modelling the orbital motion}

This Section is dedicated to the discussion of the hydrodynamic drag
on the satellite black hole. Although the two aspects are closely
inter-related, we first discuss the equations governing the
interaction between the satellite black hole and the torus and 
then describe their use in the calculation of the changes of the
orbital parameters within the adiabatic approximation.

\subsection{\label{sec:interaction}The hydrodynamic drag}

As already mentioned in Sec. \ref{sec:intro}, the hydrodynamic drag
acting on the satellite black hole can be written as the sum of a
short-range part, due to accretion, and a long-range part, due to the
deflection of the matter which is not accreted or, equivalently, to
the gravitational interaction of the satellite with the density
perturbations gravitationally induced by its own presence:
\eq\label{eq:total_rate} 
\frac{dp_{\rm
sat}^{\mu}}{d\tau}=\frac{dp^{\mu}}{d\tau}\Big\vert_{\rm
accr}+\frac{dp^{\mu}}{d\tau}\Big\vert_{\rm defl}\,, 
\eeq
where $\tau$ is the proper time of the satellite.

Accretion onto a moving black hole was studied analytically in a 
Newtonian framework by Bondi \& Hoyle~\cite{bondi}, who found the 
rest-mass accretion rate to be 
\eq \frac{dm_0}{d\tau}=  \frac{4 \pi \lambda m^2
\rho_0}{(v^2+v_s^2)^{3/2}}\;,\label{eq:bondi_rate} \eeq
where $m$ is the mass of the black hole, $v$ and $v_s$ are
respectively the velocity of the black hole with respect to the fluid
and the sound velocity, and $\lambda$ is a dimensionless constant of
the order of unity, which for a fluid with polytropic equation of
state and polytropic index $\Gamma$ has the value~\cite{BHWDNSbook}
\eq\label{eq:canonical}
\lambda= \left(\frac{1}{2}\right)^{(\Gamma+1)/[2(\Gamma-1)]}
\left(\frac{5-3\Gamma}{4}\right)^{-(5-3\Gamma)/[2(\Gamma-1)]}.
\eeq
Subsequent numerical work~\cite{petrich, font1, font2} treated instead the
problem of accretion in full General Relativity, and showed that
Eq. \eqref{eq:bondi_rate}, with $\lambda$ given by
Eq. \eqref{eq:canonical}, is correct provided that it is multiplied by
a factor $\sim5$ -- $25$ when $v$ and $v_s$ become relativistic (\textit{cf.} Table 3 of Ref.~\cite{font1}).
However, because a fit for this correction factor is, to the best of
our knowledge, not yet available, and the published data is not
sufficient for producing one, we use the Bondi accretion rate
[Eqs. \eqref{eq:bondi_rate} and \eqref{eq:canonical}], bearing in mind
that it could slightly underestimate the drag at relativistic
velocities $v$ and $v_s$.\footnote{As we will see in section \ref{sec:results}, $v$ and $v_s$ can become relativistic 
only for orbits counter-rotating with respect the torus and very close to the SMBH. For these orbits the dominant part of the hydrodynamic
drag is the long-range one, and the relativistic correction factor to the Bondi accretion rate (which is roughly $5-10$ for these orbits, as can be seen comparing the middle panel of Fig. \ref{fig:eq_circ_fig} with Table 3 of Ref.~\cite{font1}) does not change this conclusion.}  
Once the accretion rate is known, the
short-range part of the drag reads~\cite{petrich}
\eq \frac{dp^{\mu}}{d\tau}\Big\vert_{\rm accr}= h\,
\frac{dm_0}{d\tau}\, u_{\rm fluid}^\mu\;, \eeq
where we recall that $h$ is the specific enthalpy of the fluid. Note that this equation basically follows
from the conservation of the total 4-momentum of the satellite and the fluid.

The long-range drag is instead more complicated. The gravitational
interaction of a body with the density perturbations that it excites
gravitationally in the surrounding medium was first studied by
Chandrasekhar~\cite{chandra} in the case of a collisionless fluid, and
is also known as ``dynamical friction''.  Although less well
recognized, dynamical friction acts also for a body moving in a
collisional
medium~\cite{rephaeli,petrich,ostriker,sanchez,kim,my_dyn_frict}. In
particular, a satellite moving on a circular planar orbit
(\textit{e.g.,} a circular orbit around a Schwarzschild black hole or
a circular equatorial orbit around a Kerr black hole) experiences a
drag in the tangential
direction~\cite{rephaeli,petrich,ostriker,sanchez} and one in the
radial direction~\cite{kim}:
\eq\label{eq:DF_decomposed}
\frac{dp^{\mu}}{d\tau}\Big\vert_{\rm defl}=\frac{dp}{d\tau}\Big\vert_{\rm defl}^{\rm tang} \sigma^\mu+
\frac{dp}{d\tau}\Big\vert_{\rm defl}^{\rm rad} \chi^\mu\;, 
\eeq
where ${\boldsymbol \sigma}$ is a unit spacelike vector orthogonal to
${\boldsymbol u}_{\rm sat}$ and pointing in the direction of the
motion of the fluid,
\eq{\boldsymbol \sigma}=\frac{{\boldsymbol u}_{\rm fluid}-\gamma {\boldsymbol u}_{\rm
sat}}{\sqrt{\gamma^2-1}}\, \label{eq:sigma_def}\eeq
(the Lorentz factor $\gamma=-{\boldsymbol u}_{\rm fluid}\cdot{\boldsymbol u}_{\rm
sat}$ encodes the relative motion of the satellite with respect to
the fluid of the torus), and 
\eq\label{eq:chi_def} {\boldsymbol \chi}=-\frac{ u^{\rm sat}_r
{\boldsymbol u}_{\rm sat}-\sigma_r{\boldsymbol\sigma}
+\partial/\partial r}{[g_{rr}-(u^{\rm sat}_r)^2/(\gamma^2-1)]^{1/2}}\,,
\eeq
is a unit spacelike vector, orthogonal to both ${\boldsymbol u}_{\rm
sat}$ and ${\boldsymbol \sigma}$ and poiting in the radial direction.
In particular, the tangential and radial drags are given
by~\cite{kim,my_dyn_frict}
\begin{gather}
\frac{dp}{d\tau}\Big\vert_{\rm defl}^{\rm tang}=\frac{4 \pi (p+\rho) m^2\gamma^2 (1+v^2)^2}{v^2}\, I_{\rm tang}\,,\label{eq:tang_drag}\\
\frac{dp}{d\tau}\Big\vert_{\rm defl}^{\rm rad}=\frac{4 \pi (p+\rho) m^2\gamma^3 (1+v^2)^2}{v^2}\, I_{\rm rad}\,,\label{eq:rad_drag}
\end{gather}
where $I_{\rm tang}$ and $I_{\rm rad}$ are complicated integrals.
Fits to the numerically-computed steady-state\footnote{Fortunately,
the steady-state values for these integrals are reached over
timescales which are comparable with either the sound crossing-time
$r/v_s$, $r$ being the radius of the circular orbit, or with the
orbital period.} values for these integrals are given in
Ref.~\cite{kim}:

\begin{equation}\label{eq:Iphi}
  I_{\rm tang} = \left\{\begin{array}{l}
    0.7706\ln\left(\frac{1+{\cal M}}{1.0004-0.9185{\cal M}}\right)-1.4703{\cal M}, \\\mbox{for }{\cal M}<1.0\,,\\\\
     \ln[330 (r/r_{\rm min}) ({\cal M}-0.71)^{5.72}{\cal M}^{-9.58} ],\\\mbox{for }1.0\leq{\cal M}<4.4\,, \\\\
     \ln[(r/r_{\rm min})/(0.11{\cal M}+1.65)],\\ \mbox{for }{\cal M}\geq4.4\,,
  \end{array}\right.
\end{equation}
and 
\begin{equation}\label{eq:IR}
  I_{\rm rad} = \left\{\begin{array}{l}
    {\cal M}^2\ 10^{\ 3.51{\cal M}-4.22},\quad   \mbox{for }{\cal M}<1.1\,,\\\\
    0.5\  \ln\big[9.33{\cal M}^2({\cal M}^2-0.95)\big],\\ \mbox{for }1.1\leq{\cal M}<4.4\,,\\\\
    0.3\ {\cal M}^2, \quad  \mbox{for }{\cal M}\geq4.4\,,
  \end{array}\right.
\end{equation}
where $r$ is the radius of the circular orbit, $r_{\rm min}\sim 2m
(1+v^2)/v^2$ is the capture impact parameter of the satellite
black-hole, while ${\cal M}=v/v_{s}$ is the Mach number.

These fits are valid for $r\gg r_{\min}$ and are accurate within 4\%
for ${\cal M}<4.4$ and within 16\% for ${\cal M}>4.4$.  
However, the fit for $I_{\rm tang}$ does not go to zero when
${\cal M}$ goes to zero, while $I_{\rm rad}$ goes to zero only as ${\cal
M}^2$ in this limit: these behaviors would give a non-zero radial drag and a diverging 
tangential drag for $v\to0$ [\textit{cf.} Eqs. \eqref{eq:tang_drag} and
\eqref{eq:rad_drag}]. This is clearly a spurious behavior: dynamical friction
must vanish for $v=0$, since in this case the pattern of the density perturbations is spherically
symmetric around the body (as there is no preferred direction).  
However, as we will see in Sec. \ref{sec:adiabatic}, the effect of the
radial drag vanishes if one uses the \textit{adiabatic approximation}
(as it is usually done in EMRI-studies~\cite{drasco_scalar,drasco_GW,adiabatic,mino}), and
therefore this artifact of the fit \eqref{eq:IR} cannot cause
any harm in our numerical code. This is instead not the case for the
tangential drag: in order to eliminate its spurious divergence, 
we have approximated $I_{\rm tang}$ with its straight-line functional form at
low Mach numbers. Since the dynamical friction drag for straight-line
subsonic motion is given by Eq. \eqref{eq:tang_drag} with $I_{\rm
tang}=1/2 \ln[(1+{\cal M})/(1-{\cal M})]-{\cal M}\approx{\cal
M}^3/3+{\cal M}^5/5$, we can assume that $I_{\rm tang}$ is given, for
${\cal M}<0.1$, by
\eq\label{eq:addon}
I_{\rm tang}=0.9563\left(\frac{{\cal M}^3}{3}+\frac{{\cal M}^5}{5}\right)\,,
\eeq
where the factor $0.9563$ is introduced to match the above fit at
${\cal M}=0.1$.

Note that although Eq. \eqref{eq:DF_decomposed} is strictly valid only
for circular planar motion (\textit{i.e.,} in the case of a Kerr
spacetime, for circular equatorial orbits), we expect it to be a good
approximation also for generic orbits around a Kerr black
hole. Indeed, thanks to the choice of the unit vectors
$\boldsymbol{\sigma}$ and $\boldsymbol{\chi}$,
Eq. \eqref{eq:DF_decomposed} gives a tangential drag parallel the
direction of the flow and a drag in the radial direction perpendicular
to the direction of the flow.  Both of these components are expected
to be present also for generic orbits. In particular, the tangential
drag should be given approximately by Eqs. \eqref{eq:tang_drag} and
\eqref{eq:addon} if the radius $r$ appearing in Eq. \eqref{eq:Iphi} is
replaced by the \textit{semi-latus rectum} $p$ of the orbit [see
Eq. \eqref{eq:pei_defs} for the definition of this
quantity].\footnote{Note that the tangential drag given by
Eqs. \eqref{eq:tang_drag}, \eqref{eq:Iphi} and \eqref{eq:addon} is
approximately correct also for straight-line motion, if $r$ replaced
in Eq. \eqref{eq:Iphi} by $vt$ -- $t$ being the time for which the
satellite has been active~\cite{ostriker,my_dyn_frict} -- as long as
$vt$ is smaller than the size of the medium, and by a cutoff-length of
order of the size of the medium at later times. To see this, compare
Eqs. \eqref{eq:Iphi} and \eqref{eq:addon} to the functional form of
$I_{\rm tang}$ for straight-line motion, which is $I_{\rm tang}=1/2
\ln[(1+{\cal M})/(1-{\cal M})]-{\cal M}$ for subsonic motion and
$I_{\rm tang}=1/2\ln(1-1/{\cal M}^2)+\ln(vt/r_{\min})$ for supersonic
motion~\cite{ostriker,my_dyn_frict}.} Although this prescription is
not exact, the results of Ref.~\cite{kim} suggest that the relevant lengthscale in the Coulomb logarithm
appearing in the second and third lines of eq.~\eqref{eq:Iphi} should be one characterizing the orbit, rather than the size of the
medium, as commonly assumed in most of the works on dynamical friction predating Refs.~\cite{ostriker,kim} (see the introduction of
Ref.~\cite{kim} and references therein for more details about this point). Of course, 
this lengthscale could be different from the semi-latus rectum of the orbit, but different choices for it would
have only a slight impact on the results because of the logarithmic dependence.  

The extrapolation of the
radial drag given by Eqs.  \eqref{eq:rad_drag} and \eqref{eq:IR} from
circular planar to generic orbits is instead a bit more problematic,
although one expects it to be a good approximation at least for orbits
with small eccentricities and small inclinations with respect to the
equatorial plane.  At any rate, as we have mentioned earlier, in
Sec. \ref{sec:adiabatic} we will show that the effect of this radial
drag on the orbital evolution averages to zero when adopting the
\textit{adiabatic approximation}. (Note that this agrees with
Ref.~\cite{kim}, which found that the effect of the radial drag on the
orbital evolution was subdominant with respect to that of the
tangential drag.)  Nevertheless, a non-zero effect may still be
present in cases in which the adiabatic approximation is not valid
(\textit{i.e.} if the hydrodynamic drag acts on a timescale comparable
to the orbital period), or possibly even in the adiabatic
approximation if more rigorous expressions for the radial drag should
be derived in the future.

The rate of change of the mass of the satellite with respect to the
coordinate time $t$ follows immediately from $ p_{\rm
sat}^{\mu}\,p^{\rm sat}_{\mu}=-m^2$: denoting the derivative with
respect to $t$ with an overdot, we have
\eq\label{eq:mdot}
\dot m=-\frac{u^{\rm sat}_{\mu}}{u^t_{\rm sat}}
\frac{d p^\mu_{\rm sat}}{d\tau}=-\frac{u^{\rm sat}_{\mu}}{{u^t_{\rm sat}}}
\frac{d p^\mu_{\rm accr}}{d\tau}=
\frac{h\,\gamma}{u_{\rm sat}^t}\frac{d m_0}{d\tau}\;.
\eeq

It is well-known~\cite{carter} that Kerr geodesics can be labeled, up to initial conditions, 
by three constants of motion, the dimensionless energy $\tilde{E}$  and
the angular momentum $\tilde{L}_z$ as measured by an observer at infinity, 
\begin{equation}\label{eq:EL_defs}
\tilde{E}=-u^{\rm sat}_t\,,\quad \tilde{L}_z=u^{\rm sat}_\phi/M\,,
\end{equation}
and the dimensionless Carter constant~\cite{carter} $\tilde{Q}$,
\begin{equation}\label{eq:Qdef}
\tilde{Q}=\left(\frac{u^{\rm sat}_\theta}{M}\right)^2 +
 \tilde{a}^2 \cos^2 \theta (1 -\tilde{E}^2) +  \cot^2 \theta \tilde{L}_z^2
\end{equation}
where $\tilde{a}=a/M$. We will now derive expression for the rates of
change of these quantities.

To this purpose, let us first introduce the tetrad $({\boldsymbol
u}_{\rm sat}, {\boldsymbol e}_{1}= {\boldsymbol \sigma}, {\boldsymbol
e}_{2}= {\boldsymbol \chi}, {\boldsymbol e}_{3})$ based in the
position of the satellite and write the change in the 4-velocity due
to accretion and deflection of the flow as
\eq\label{eq:delta_u1}
\delta u_{\rm sat}^\mu=
\delta u_{\rm sat}^{(t)}u_{\rm sat}^\mu+\delta u^{(i)}_{\rm sat}e_{(i)}^\mu\;,
\eeq
where $\delta u_{\rm sat}^{(t)}$ and $\delta u_{\rm sat}^{(i)}$ are
the components with respect to the tetrad. In particular, perturbing
$-(u_{\rm sat}^{(t)})^2+\delta_{ij}u_{\rm sat}^{(i)}u_{\rm
sat}^{(j)}=-1$ to first order one easily gets $-u_{\rm
sat}^{(t)}\delta u_{\rm sat}^{(t)}+\delta_{ij}u_{\rm sat}^{(i)}\delta
u_{\rm sat}^{(j)}=0$, and using then the fact that $u_{\rm
sat}^{(i)}=0$ to zeroth order, one obtains $\delta u_{\rm
sat}^{(t)}=0$. Using now $\delta u_{\rm sat}^{(i)}=\delta p_{\rm
sat}^{(i)}/m$, ${\boldsymbol e}_{(i)}\cdot {\boldsymbol u}_{\rm
sat}=0$ and ${\boldsymbol e}_{(i)}\cdot {\boldsymbol
e}_{(j)}=\delta_{ij}$ ($i=1,2,3$), Eq. \eqref{eq:delta_u1} becomes
\begin{multline}\label{eq:delta_u2}
\delta u_{\rm sat}^\mu=\left(\frac{\delta m_0\,h}{m} u_{\rm fluid}^\nu\,
\sigma_\nu+\frac{\delta p_{\rm defl}^{\rm tang}}{m}\right) \sigma^\mu+\frac{\delta p_{\rm defl}^{\rm rad}}{m}\chi^\mu
\\=\left(\frac{\delta m_0\,h}{m}+\frac{\delta p_{\rm defl}^{\rm tang}}{m\sqrt{\gamma^2-1}}\right)
\left(u_{\rm fluid}^\mu-\gamma u_{\rm sat}^\mu\right)+\frac{\delta p_{\rm defl}^{\rm rad}}{m}\chi^\mu\;.
\end{multline}
Using now Eqs. \eqref{eq:u}, \eqref{eq:EL_defs} and \eqref{eq:delta_u2},
we immediately obtain
\begin{gather}
\frac{\dot {\tilde{E}}}{\tilde{E}}=
\left(\frac{\dot m_0 \,h}{m}+\frac{\dot p_{\rm defl}^{\rm tang}}{m\sqrt{\gamma^2-1}}\right)
\left(\frac{U}{\tilde{E}}-\gamma\right)-\frac{\dot p_{\rm defl}^{\rm rad}}{m\tilde{E}}\chi_t\;,\label{eq:Edot_bondi}\\
\frac{\dot {\tilde{L}}_z}{{\tilde{L}}_z}= \left(\frac{\dot
m_0 \,h}{m}+\frac{\dot p_{\rm defl}^{\rm tang}}{m\sqrt{\gamma^2-1}}\right)
\left(\frac{\ell\,U}{M\tilde{L}_z}-\gamma\right)+\frac{\dot p_{\rm defl}^{\rm rad}}{mM\tilde{L}_z}\chi_\phi\;.\label{eq:Ldot_bondi}
\end{gather}
In order to calculate instead the rate of change of the
dimensionless Carter constant $\tilde Q$, 
let us note that from Eq. \eqref{eq:delta_u2} it
follows that the variation of $u_\theta$ in a short time interval $\delta t$ due to accretion and
deflection of the flow is
\eq\label{eq:u_theta}
\delta u^{\rm sat}_\theta=\left[-\gamma \left(\frac{\dot m_0 \,h}{m}+
\frac{\dot p^{\rm tang}_{\rm defl}}{m\sqrt{\gamma^2-1}}\right) u^{\rm sat}_\theta+\frac{\dot p_{\rm defl}^{\rm rad}}{m}\chi_\theta\right]\delta t\;.
\eeq
We can then write ${\dot u}^{\rm sat}_\theta$ as the sum of a term
coming from the gravitational evolution (\textit{i.e.,} the geodesic
equation) and one coming from collisions with the surrounding gas:
 \begin{multline}
 {\dot u}^{\rm sat}_\theta=\Gamma_{\theta\nu}^\mu u^{\rm sat}_\mu {\dot x}_{\rm sat}^\nu
 \\-\gamma\left(\frac{\dot m_0 \,h}{m}+\frac{\dot p^{\rm tang}_{\rm defl}}{m\sqrt{\gamma^2-1}}\right)u^{\rm sat}_\theta+\frac{\dot p_{\rm defl}^{\rm rad}}{m}\chi_\theta\;.
 \end{multline}
 The evolution of $\tilde Q$ therefore follows from Eq. \eqref{eq:Qdef}:
 \begin{align}
 &{\dot {\tilde Q}}=\,\frac{\partial{\tilde Q}}{\partial \theta}{\dot \theta}_{\rm sat}
 +\frac{\partial{\tilde Q}}{\partial u_\theta^{\rm sat}}\Gamma_{\theta\nu}^\mu u_\mu^{\rm sat} {\dot x}^\nu_{\rm sat}
 +\frac{\partial{\tilde Q}}{\partial {\tilde E}}{\dot {\tilde E}}
 +\frac{\partial{\tilde Q}}{\partial {\tilde L}_z}{\dot {\tilde L}_z}
\nonumber
 \\&-\frac{\partial{\tilde Q}}{\partial u_\theta^{\rm sat}}\,\gamma
 \left(\frac{\dot m_0 \,h}{m}+\frac{\dot p^{\rm tang}_{\rm defl}}{m\sqrt{\gamma^2-1}}\right)u^{\rm sat}_\theta 
+\frac{\partial{\tilde Q}}{\partial u_\theta^{\rm sat}}\,\frac{\dot p_{\rm defl}^{\rm rad}}{m}\chi_\theta 
\nonumber\\ &=
\frac{\partial{\tilde Q}}{\partial \tilde{E}}{\dot
{\tilde E}}+ \frac{\partial{\tilde Q}}{\partial {\tilde
L}_z}{\dot {\tilde L}_z}
-\frac{\partial{\tilde Q}}{\partial u_\theta^{\rm
sat}}\,\gamma \left(\frac{\dot m_0 \,h}{m}+\frac{\dot p_{\rm
defl}^{\rm tang}}{m\sqrt{\gamma^2-1}}\right)u^{\rm sat}_\theta
\nonumber\\ 
&+\frac{\partial{\tilde Q}}{\partial u_\theta^{\rm sat}}\,\frac{\dot p_{\rm defl}^{\rm rad}}{m}\chi_\theta ,\label{eq:Qdot_bondi}
 \end{align}
where the partial derivatives are meant to be calculated with
Eq. \eqref{eq:Qdef}. Note that the first and the second term of the
first line cancel out because $\tilde Q$ is conserved for geodesic
motion.

A useful alternative form for the evolution rate of $\tilde{Q}$ can be obtained by 
rewriting Eq. \eqref{eq:Qdef} using the normalization condition $\boldsymbol{u}^{\rm sat}\cdot\boldsymbol{u}^{\rm sat}=-1$:
\begin{multline}\label{eq:Qdef2}
\tilde{Q}=\tilde{\Delta}^{-1}\left[ {\tilde{E}(\tilde{r}^2+\tilde{a}^2)-\tilde{a} \tilde{L} } \right]^2\\-(\tilde{L}
-\tilde{a}\tilde{E})^2-\tilde{r}^2- \tilde{\Delta} {(u_r^{\rm sat})}^2\;,
\end{multline}
where $\tilde{r}= r/M$ and $\tilde{\Delta}= \Delta/M^2$.  Proceeding
as above and in particular using the fact that
 \begin{multline}
 {\dot u}^{\rm sat}_r=\Gamma_{r\nu}^\mu u^{\rm sat}_\mu {\dot x}_{\rm sat}^\nu\\
 -\gamma\left(\frac{\dot m_0 \,h}{m}+\frac{\dot p^{\rm tang}_{\rm defl}}{m\sqrt{\gamma^2-1}}\right)u^{\rm sat}_r
 +\frac{\dot p_{\rm defl}^{\rm rad}}{m}\chi_r\;
 \end{multline}
[from Eqs. \eqref{eq:u} and \eqref{eq:delta_u2}], one easily gets
 \begin{align}
 &{\dot {\tilde Q}}=\,\frac{\partial{\tilde Q}}{\partial r}{\dot r}_{\rm sat}
 +\frac{\partial{\tilde Q}}{\partial u_r^{\rm sat}}\Gamma_{r\nu}^\mu u_\mu^{\rm sat} {\dot x}^\nu_{\rm sat}
 +\frac{\partial{\tilde Q}}{\partial {\tilde E}}{\dot {\tilde E}}
 +\frac{\partial{\tilde Q}}{\partial {\tilde L}_z}{\dot {\tilde L}_z}
\nonumber
 \\&-\frac{\partial{\tilde Q}}{\partial u_r^{\rm sat}}\,\gamma
 \left(\frac{\dot m_0 \,h}{m}+\frac{\dot p^{\rm tang}_{\rm defl}}{m\sqrt{\gamma^2-1}}\right)u^{\rm sat}_r
+\frac{\partial{\tilde Q}}{\partial u_r^{\rm sat}}\,\frac{\dot p_{\rm defl}^{\rm rad}}{m}\chi_r
\nonumber\\ &=\frac{\partial{\tilde Q}}{\partial \tilde{E}}{\dot
{\tilde E}}+ \frac{\partial{\tilde Q}}{\partial {\tilde
L}_z}{\dot {\tilde L}_z}-\frac{\partial{\tilde Q}}{\partial u_r^{\rm
sat}}\,\gamma \left(\frac{\dot m_0 \,h}{m}+\frac{\dot p_{\rm
defl}^{\rm tang}}{m\sqrt{\gamma^2-1}}\right)u^{\rm sat}_r\nonumber\\
&+\frac{\partial{\tilde Q}}{\partial u_r^{\rm sat}}\,\frac{\dot p_{\rm defl}^{\rm rad}}{m}\chi_r
\;,\label{eq:Qdot_bondi2}
 \end{align}
where the partial derivatives are now calculated with
Eq. \eqref{eq:Qdef2}. Note that for circular orbits
Eq. \eqref{eq:Qdot_bondi2} becomes
\eq\label{eq:circ2circ}
{\dot {\tilde Q}}=\frac{\partial{\tilde Q}}{\partial {\tilde E}}{\dot {\tilde E}}
 +\frac{\partial{\tilde Q}}{\partial {\tilde L}_z}{\dot {\tilde L}_z}
\eeq
[use Eq. \eqref{eq:Qdef2} and the fact that $u_r^{\rm sat}=0$ for circular orbits].
This condition ensures\footnote{Note in particular that the proof
presented in Ref.~\cite{kennefick_ori}, which was concerned mainly with radiation reaction, applies also to the case of the hydrodynamic drag. Note also that the resonance condition which was found in Ref.~\cite{kennefick_ori} as the only possible case that could give rise to
a non-circular evolution for an initially circular orbit is never satisfied in a Kerr spacetime~\cite{ryan2}.}
that circular orbits keep circular under the hydrodynamic drag and in the adiabatic approximation, as it happens for 
radiation reaction.

Finally, let us note that the rates of change of ${\tilde E}$,
${\tilde L}_z$ and ${\tilde Q}$ [Eqs. \eqref{eq:Edot_bondi},
\eqref{eq:Ldot_bondi}, \eqref{eq:Qdot_bondi} and
\eqref{eq:Qdot_bondi2}] go smoothly to zero as the velocity of the
satellite relative to the fluid goes to zero.  This is easy to check
using the fact that, when $v$ approaches zero, $\dot{p}_{\rm defl}^{\rm tang}={\cal O}(v)$ 
[\textit{cf.} Eqs.~\eqref{eq:tang_drag} and~\eqref{eq:addon}], $\dot{p}_{\rm defl}^{\rm rad}\to0$,
$\gamma^2-1={\cal O}(v^2)$,  $u_r={\cal O}(v)$, $u_\theta={\cal O}(v)$,
$\ell U- M\tilde{L}_z={\cal O}(v)$ and $U-\tilde{E}={\cal O}(v^2)$,
and using the fact that $\boldsymbol{\chi}$ keeps finite in this limit 
[in particular, from Eqs. \eqref{eq:sigma_def} and \eqref{eq:chi_def} 
it follows $\chi_t={\cal O}(v)$, $\chi_\phi={\cal O}(1)$, $\chi_\theta={\cal
O}(1)$ and $\chi_r={\cal O}(1)$].  Note that this is indeed the result
that one would expect.  First of all, a body comoving with the fluid
clearly does not experience any dynamical friction and the only active
mechanism is accretion.  The body then accretes mass and consequently
energy and angular momentum (because the fluid carries a specific
energy and a specific angular momentum).  However, the
\textit{dimensionless} constants of motion ${\tilde E}$, ${\tilde
L}_z$ and ${\tilde Q}$ entering the geodesic equation cannot change
because of the weak equivalence principle.  Pictorially, one may
think of a satellite comoving with a gaseous medium.
Consider a sphere centered in the satellite, with radius small enough to ensure 
that the gas contained in the sphere has approximately the same velocity as the satellite.
Suppose now that all the gas in this sphere is accreted by the satellite.
The velocity of the satellite will clearly be unaffected, because of the 
conservation of momentum: for the weak equivalence principle this is enough to ensure that the orbit of the satellite will be unaffected,
in spite of its increased mass.

\subsection{\label{sec:adiabatic}The adiabatic approximation}

At the heart of our approach is the calculation of the changes of the
orbital parameters experienced by Kerr
geodesics as a result of the hydrodynamic drag, and their comparison
with the corresponding changes introduced by radiation reaction. To
this purpose, let us recall that up to initial conditions Kerr
geodesics can be labeled by a set of three parameters, the
\textit{semi-latus rectum} $p$, the \textit{eccentricity} $e$ and the
\textit{inclination angle} $\theta_{\rm inc}$. These are just a
remapping of the energy, angular momentum and Carter constant
introduced in Sec. \ref{sec:interaction}, and are defined as
\cite{schmidt}
\begin{align}\label{eq:pei_defs}
&p=\frac{2 r_{\rm a} r_{\rm p}}{r_{\rm a} +r_{\rm p}}\,,\quad e=\frac{r_{\rm a} -r_{\rm p}}{r_{\rm a} +r_{\rm p}}\,,\nonumber\\
&\theta_{\rm inc} = \frac{\pi}{2}-D\,\theta_{\min}\;,
\end{align}
where $r_{\rm a}$ and $r_{\rm p}$ are the apastron and periastron
coordinate radii, $\theta_{\min}$ is the minimum polar angle $\theta$
reached during the orbital motion and $D=1$ for orbits corotating with
the SMBH whereas $D=-1$ for orbits counter-rotating with
respect to it. Note that in the weak-field limit $p$ and $e$
correspond exactly to the semi-latus rectum and eccentricity used to
describe orbits in Newtonian gravity, and that $\theta_{\rm inc}$ goes
from $\theta_{\rm inc} = 0$ for equatorial orbits corotating with the
black hole to $\theta_{\rm inc} = 180$ degrees for equatorial orbits
counter-rotating with respect to the black hole, passing through
$\theta_{\rm inc} = 90$ degrees for polar orbits.

In order to fix the initial conditions of a geodesic, let us first
parametrize it with the Carter time $\lambda$, which is related to
the proper time by~\cite{carter}
\eq\label{eq:lambda_def} \frac{d\tau}{d\lambda}=\Sigma\;.  \eeq
This is a very useful choice because it makes the geodesic equation separable~\cite{carter}: 
\begin{align}
&\left(\frac{dr}{d\lambda}\right)^2 = V_r(r), &
&\frac{dt}{d\lambda} = V_t(r,\theta),&
\nonumber \\
&\left(\frac{d\theta}{d\lambda}\right)^2 = V_\theta(\theta), &
&\frac{d\phi}{d\lambda} = V_\phi(r,\theta)\;,&
 \label{geodesics}
\end{align}
with 
%
% \begin{subequations} \label{detailed geodesics}
\begin{align}
&V_t(r,\theta)/M^2=\nonumber
 \\& \tilde{E} \left[ \frac{(\tilde{r}^2+\tilde{a}^2)^2}{\tilde{\Delta}} - \tilde{a}^2\sin^2\theta \right]
   + \tilde{a}\tilde{L}_z \left( 1 - \frac{\tilde{r}^2+\tilde{a}^2}{\tilde{\Delta}} \right),\nonumber\\\label{tdot} 
\\
&V_r(r)/M^4=\nonumber\\
  &\left[ \tilde{E}(\tilde{r}^2+\tilde{a}^2) - \tilde{a} \tilde{L}_z \right]^2
  - \tilde{\Delta}\left[\tilde{r}^2 + (\tilde{L}_z - \tilde{a} \tilde{E})^2 + \tilde{Q}\right],\nonumber\\\label{rdot}
\end{align}
\begin{align}
&V_\theta(\theta)/M^2  =\tilde{Q} - \tilde{L}_z^2 \cot^2\theta - \tilde{a}^2(1 - \tilde{E}^2)\cos^2\theta,\nonumber\\
\label{thetadot}\\
&V_\phi(r,\theta)/M
  = \tilde{L}_z \csc^2\theta + \tilde{a}\tilde{E}\left(\frac{\tilde{r}^2+\tilde{a}^2}{\tilde{\Delta}} - 1\right) - \frac{\tilde{a}^2\tilde{L}_z}{\tilde{\Delta}}.\nonumber\\
\label{phidot}
\end{align}
% \end{subequations}
%
This means, in particular, that the $r$- and $\theta$-motions are
periodic in $\lambda$.  The initial conditions of the geodesic can
then be characterized by the values $t_0$ and $\phi_0$ of the
coordinates $t$ and $\phi$ when $\lambda=0$, the value $\lambda_{r0}$
of the Carter time nearest to $\lambda=0$ at which
$r(\lambda_{r0})=r_{\rm p}$, and the value $\lambda_{\theta0}$ of the
Carter time nearest to $\lambda=0$ at which
$\theta(\lambda_{\theta0})=\theta_{\min}$~\cite{drasco_scalar}.

Let us fix the geodesic under consideration by choosing the parameters
$p$, $e$ and $\theta_{\rm inc}$ so as to obtain a bound and stable
orbit (see Ref.~\cite{schmidt} for details) and by choosing the
initial conditions as described above. One could in principle use
Eqs. \eqref{eq:Edot_bondi}, \eqref{eq:Ldot_bondi} and
\eqref{eq:Qdot_bondi} [or \eqref{eq:Qdot_bondi2}] to compute the rates
of change of $\tilde E$, $\tilde{L}_z$ and $\tilde{Q}$ due to the
hydrodynamic drag as a function the Carter time $\lambda$.  However,
because the timescale of the orbital evolution due to the interaction
with the torus is much longer than the orbital period, we can apply
the adiabatic approximation and compute instead the averages of ${\dot
{\tilde E}}$, ${\dot {\tilde L}_z}$ and ${\dot {\tilde Q}}$ over times
much longer than the orbital periods.  This approximation is routinely
adopted when studying the effect of radiation reaction on EMRIs
\cite{drasco_scalar,drasco_GW,adiabatic,mino}, and it is easy to
implement when one considers instead the effect of the hydrodynamic
drag. Denoting respectively with $\langle\phantom{a}\rangle_t$ and
$\langle\phantom{a}\rangle_\lambda$ the average over an infinite
coordinate time and the average over an infinite Carter time, we can
write~\cite{drasco_scalar}
\begin{equation}
\langle {\dot \Psi} \rangle_t=\frac{\langle {d{\Psi}}/d\lambda
  \rangle_\lambda}{\langle dt/d\lambda
  \rangle_\lambda}\label{eq:av1}\,,
\end{equation}
where $\Psi$ is a place-holder for either ${{\tilde E}}, {{\tilde
L}_z}$ or ${{\tilde Q}}$.

Using now Eq. \eqref{eq:u} (with the assumption that the torus is
symmetric with respect to the equatorial plane) in
Eqs. \eqref{eq:Edot_bondi}, \eqref{eq:Ldot_bondi} and
\eqref{eq:Qdot_bondi} [or \eqref{eq:Qdot_bondi2}], it is easy to show
that ${d{\tilde E}}/d\lambda$, ${d {\tilde L}_z}/d\lambda$ and ${d
{\tilde Q}}/d\lambda$ depend, once fixed $\tilde E$, ${\tilde L}_z$
and $\tilde Q$, only on the $r$ and $\cos^2\theta$ of the geodesic
under consideration -- \textit{i.e.,} $r=r(\lambda,\lambda_{r0})$ and
$\cos^2\theta=\cos^2\theta(\lambda,\lambda_{\theta0})$ -- and on the
sign of $u^{\rm sat}_r$, which we will denote by $\epsilon_r$. [The
dependence on this sign arises because of the terms due to the radial
drag, as the quantity $u^{\rm sat}_r$ appearing in the definitions of
$\boldsymbol{\sigma}$ and $\boldsymbol{\chi}$ can be expressed in
terms of $r$ and $\cos^2\theta$ using
Eqs. \eqref{eq:lambda_def}-\eqref{phidot} only up to such a sign.]
Similarly, $dt/d\lambda$ is given by the geodesic equation
\eqref{geodesics} and depends, once fixed $\tilde E$, ${\tilde L}_z$
and $\tilde Q$, only on the $r$ and $\cos^2\theta$ of the geodesic
[\textit{cf.} Eq.~\eqref{tdot}].

Using now the fact that the $r$- and $\theta$-motions are periodic
when expressed in the Carter time, we can expand the functions
$d{\tilde E}/d\lambda$, $d{\tilde L}_z/d\lambda$ and $d{\tilde
Q}/d\lambda$ and $dt/d\lambda$ appearing in Eq. \eqref{eq:av1} in a
Fourier series. Noting that the oscillating terms average out, one can
then write these equations using only averages of these functions over
the $r$- and $\theta$-periods. More precisely, writing the $r$- and
$\theta$-motions as
$r(\lambda,\lambda_{r0})=\hat{r}(\lambda-\lambda_{r0})$ and
$\theta(\lambda,\lambda_{\theta0})=\hat{\theta}(\lambda-\lambda_{\theta0})$
(where we have denoted with a ``hat'' a fiducial geodesic having the
same $\tilde E$, $\tilde{L}_z$ and $\tilde{Q}$ as the geodesic under
consideration and $\lambda_{r0}=\lambda_{\theta0}=0$), using the fact
that ${d{\tilde E}/d\lambda}$, ${d{\tilde L}_z/d\lambda}$ and
${d{\tilde Q}/d\lambda}$ depend on $r$, $\cos^2\theta$ and
$\epsilon_r$, and using the fact that ${d{t}/d\lambda}$ depends only
on $r$ and $\cos^2\theta$, we can easily write
Eq.~\eqref{eq:av1} as~\cite{drasco_scalar}
\begin{equation}
 \!\!\!\langle {\dot \Psi}
 \rangle_t\!=\!\label{eq:Eflux}\frac{\int_{0}^{\Lambda_r}d\lambda_r
 \int_{0}^{\Lambda_\theta/4}  d\lambda_\theta\, d{\Psi}/d\lambda
 (\hat{r}(\lambda_r),\cos^2\hat{\theta}(\lambda_\theta),\epsilon_r)}
 {\int_{0}^{\Lambda_r}d\lambda_r
 \int_{0}^{\Lambda_\theta/4}  d\lambda_\theta \,V_t
 (\hat{r}(\lambda_r),\cos^2\hat{\theta}(\lambda_\theta))}\,,
\end{equation}
where $\Psi$ is again a place-holder for either ${{\tilde E}},
{{\tilde L}_z}$ or ${{\tilde Q}}$. Note that here $\Lambda_r$ and
$\Lambda_\theta$ are the $r$- and $\theta$-periods and that $d{\tilde
E}/d\lambda$, $d{\tilde L}_z/d\lambda$ and $d{\tilde Q}/d\lambda$ and
$dt/d\lambda$ are expressed using Eqs. \eqref{eq:Edot_bondi},
\eqref{eq:Ldot_bondi}, \eqref{eq:Qdot_bondi} [or
\eqref{eq:Qdot_bondi2}], \eqref{geodesics} and
\eqref{tdot}-\eqref{phidot}.
Using now the definitions of $\boldsymbol{\sigma}$ and
$\boldsymbol{\chi}$ [Eqs. \eqref{eq:sigma_def} and
\eqref{eq:chi_def}], it is easy to check that the changes of $\tilde
E$, ${\tilde L}_z$ and $\tilde Q$ arising from the radial drag average
out in the above equation because of the presence of the sign
$\epsilon_r$ (in particular $\chi_t,\chi_\phi,\chi_\theta\propto
\epsilon_r$).  As a result, one can assume $\dot{p}_{\rm defl}^{\rm
rad}=0$ \textit{ab initio} when computing Eq.~\eqref{eq:Eflux} and
benefit from another small simplification since, as we have already
mentioned, $d{\tilde E}/d\lambda$, $d{\tilde L}_z/d\lambda$ and
$d{\tilde Q}/d\lambda$ would depend only on $r$ and $\cos^2\theta$ if
it were not for the radial drag, which brings in the dependence on
$\epsilon_r$. With this assumption, all of the integrals appearing in
\eqref{eq:Eflux} can therefore be performed over
$\lambda_r\in[0,\Lambda_r/2]$ rather than over
$\lambda_r\in[0,\Lambda_r]$. Finally, note also that the rates of
change \eqref{eq:Eflux} do not depend on the initial conditions
$\lambda_{r0}$ and $\lambda_{\theta0}$ of the geodesic.

In order to reduce Eq. \eqref{eq:Eflux} to a form suitable for
numerical integration, we can express our fiducial geodesic with the
phase variables $\psi$ and $\chi$, defined
by~\cite{schmidt,drasco_GW,drasco_functionals}
\begin{gather}
\hat{r}(\psi)=\frac{p}{1+e \cos\psi}\,,\\
\cos\hat{\theta}(\chi) =z_{-}\cos\chi\,.
\end{gather}
Note that $\psi$ and $\chi$ change by $2\pi$ during respectively an
$r$- and a $\theta$-period.
Inserting then these definitions into the geodesic equation \eqref{geodesics} one gets~\cite{schmidt,drasco_GW,drasco_functionals}
\begin{gather}
\frac{d\psi}{d\lambda}=\frac{p}{1-e^2}\sqrt{J(\psi)}\;,\\
\frac{d\chi}{d\lambda}=M\sqrt{\beta(z_{+}-z_{-}\cos^2\chi)}\;,
\end{gather}
where
\begin{align}
&J(\psi)= (1-\tilde{E}^2)(1-e^2)\nonumber\\&+2\left(1-\tilde{E}^2-\frac{1-e^2}{\tilde{p}}\right)(1+e\cos\psi)+
(1+e\cos\psi)^2\times\nonumber\\&\times\left[(1-\tilde{E}^2)\frac{3+e^2}{1-e^2} -\frac{4}{\tilde{p}}+\frac{1-e^2}{\tilde{p}^2}(\beta+\tilde{L}_z^2+\tilde{Q})\right]\;,\\
&\beta= \tilde{a}^2 (1-\tilde{E}^2)\;,\\
&z_{+}=\frac{\tilde{Q}+\tilde{L}_z^2+\beta+\sqrt{(\tilde{Q}+\tilde{L}_z^2+\beta)^2-4 \beta\tilde{Q}}}{2 \beta}\,,
\end{align}
with $\tilde{p}=p/M$.  Note that ${d\psi}/{d\lambda}$ and
${d\chi}/{d\lambda}$, differently from ${dr}/{d\lambda}$ and
${d\theta}/{d\lambda}$, are non-zero at the inversion points of the
$r$- and $\theta$-motions, making $\psi$ and $\chi$ very useful for
numerical integration.

Changing the integration variables $\lambda_r$ and $\lambda_\theta$ to
$\psi$ and $\chi$, Eq.~\eqref{eq:Eflux} becomes
\begin{multline}
\langle {\dot {\Psi}} \rangle_t=\label{eq:Eflux2}\\
\int_{0}^{\pi}d\psi\int_{0}^{\pi/2} \!\!d\chi 
\frac{ d{\Psi}/d\lambda\vert_{\dot{p}_{\rm
defl}^{\rm rad}=0}
(\hat{r}(\psi),
\cos^2\hat{\theta}(\chi))(1-e^2)}{p\sqrt{J(\psi)\beta(z_{+}-z_{-}\cos^2\chi)}}\,
\times\\\times\left[
\int_{0}^{\pi}d\psi\int_{0}^{\pi/2} \!\!d\chi 
\!\frac{
V_t (\hat{r}(\psi),\cos^2\hat{\theta}(\chi))
(1-e^2)}{p\sqrt{J(\psi)\beta(z_{+}-z_{-}\cos^2\chi)}} \right]^{-1}\,.
\end{multline}
Note that the two-dimensional integrals involved in these expressions
can be easily computed numerically (\textit{e.g.,} iterating Romberg's
method~\cite{nr}) once fixed the orbital parameters $p$, $e$,
$\theta_{\rm inc}$ of the geodesic under consideration.

\begin{table*}[tbh]
\begin{tabular}{|c|c|c|c|c|c|c|c|c|c|c|c|c|c|c|c|c|}
\hline
 \T \B  Model & $a$ & $M_{\rm t}/M$ & $r_{\rm out}/M$ & $\kappa$  (CGS) & $\ell/M$ & $r_{\rm in}/M$ &
 $r_{\rm center}/M$ & $\rho_{\rm center}$ (g/cm$^3$) & $\rho_{0\rm avg}$ (g/cm$^3$) \\
\hline
\T \B A1 &     0.900&     0.100& 1.000$\times$10$^{5}$& 4.198$\times$10$^{22}$&     2.6324500536&     1.73246&     3.60963& 4.060$\times$10$^{-5}$& 1.475$\times$10$^{-11}$\\
\T \B A2 &    -0.900&     0.100& 1.000$\times$10$^{5}$& 4.189$\times$10$^{22}$&     4.7567317819&     5.65700&    15.58890& 3.992$\times$10$^{-6}$& 1.476$\times$10$^{-11}$\\
\hline
\T \B B1 &     0.998&     0.100& 1.000$\times$10$^{5}$& 4.200$\times$10$^{22}$&     2.0894422310&     1.09144&     1.56484& 1.868$\times$10$^{-4}$& 1.474$\times$10$^{-11}$\\
\T \B B2 &     0.500&     0.100& 1.000$\times$10$^{5}$& 4.195$\times$10$^{22}$&     3.4141929560&     2.91425&     7.16458& 1.331$\times$10$^{-5}$& 1.475$\times$10$^{-11}$\\
\T \B B3 &     0.000&     0.100& 1.000$\times$10$^{5}$& 4.192$\times$10$^{22}$&     3.9999599993&     4.00008&    10.47174& 7.355$\times$10$^{-6}$& 1.475$\times$10$^{-11}$\\
\T \B B4 &    -0.500&     0.100& 1.000$\times$10$^{5}$& 4.190$\times$10$^{22}$&     4.4494291313&     4.94962&    13.39547& 5.034$\times$10$^{-6}$& 1.475$\times$10$^{-11}$\\
\T \B B5 &    -0.998&     0.100& 1.000$\times$10$^{5}$& 4.188$\times$10$^{22}$&     4.8269302324&     5.82521&    16.11218& 3.796$\times$10$^{-6}$& 1.476$\times$10$^{-11}$\\
\hline
\T \B C1 &     0.900&     0.100& 1.000$\times$10$^{3}$& 3.997$\times$10$^{20}$&     2.6319080229&     1.73318&     3.60622& 4.323$\times$10$^{-2}$& 1.507$\times$10$^{-5}$\\
\T \B C2 &    -0.900&     0.100& 1.000$\times$10$^{3}$& 3.607$\times$10$^{20}$&     4.7490561067&     5.67540&    15.49760& 4.775$\times$10$^{-3}$& 1.573$\times$10$^{-5}$\\
\T \B C3 &     0.900&     0.100& 1.000$\times$10$^{4}$& 4.170$\times$10$^{21}$&     2.6324007478&     1.73253&     3.60932& 1.295$\times$10$^{-3}$& 1.479$\times$10$^{-8}$\\
\T \B C4 &    -0.900&     0.100& 1.000$\times$10$^{4}$& 4.103$\times$10$^{21}$&     4.7560304461&     5.65866&    15.58057& 1.297$\times$10$^{-4}$& 1.489$\times$10$^{-8}$\\
\T \B C5 &     0.900&     0.100& 1.000$\times$10$^{6}$& 4.201$\times$10$^{23}$&     2.6324549842&     1.73246&     3.60967& 1.283$\times$10$^{-6}$& 1.475$\times$10$^{-14}$\\
\T \B C6 &    -0.900&     0.100& 1.000$\times$10$^{6}$& 4.199$\times$10$^{23}$&     4.7568019526&     5.65683&    15.58974& 1.258$\times$10$^{-7}$& 1.475$\times$10$^{-14}$\\
\hline
\end{tabular}
\caption{Models analyzed in this paper: all of them have $M=10^6
M_{\odot}$, $\Gamma=5/3$ and are filling exactly their outermost
closed equipotential surface (\textit{i.e.,} they have $\Delta W=0$).
All the parameters are defined in Sec. \ref{sec:torus}, except the
average rest-mass density $\rho_{0 \rm avg}= M_{{\rm t} 0}/V$, where
$V=\int_{\rho_0>0}\sqrt{-g}\,d^3x$. Note that the specific angular
momentum of the torus needs to be tuned with high accuracy in order to
obtain large outer radii such as those considered in these models, and
for this reason we report $\ell/M$ with 10 decimal digits.
\label{table:models}}
\end{table*}

\section{\label{sec:results}Results}
\begin{figure*}
\includegraphics{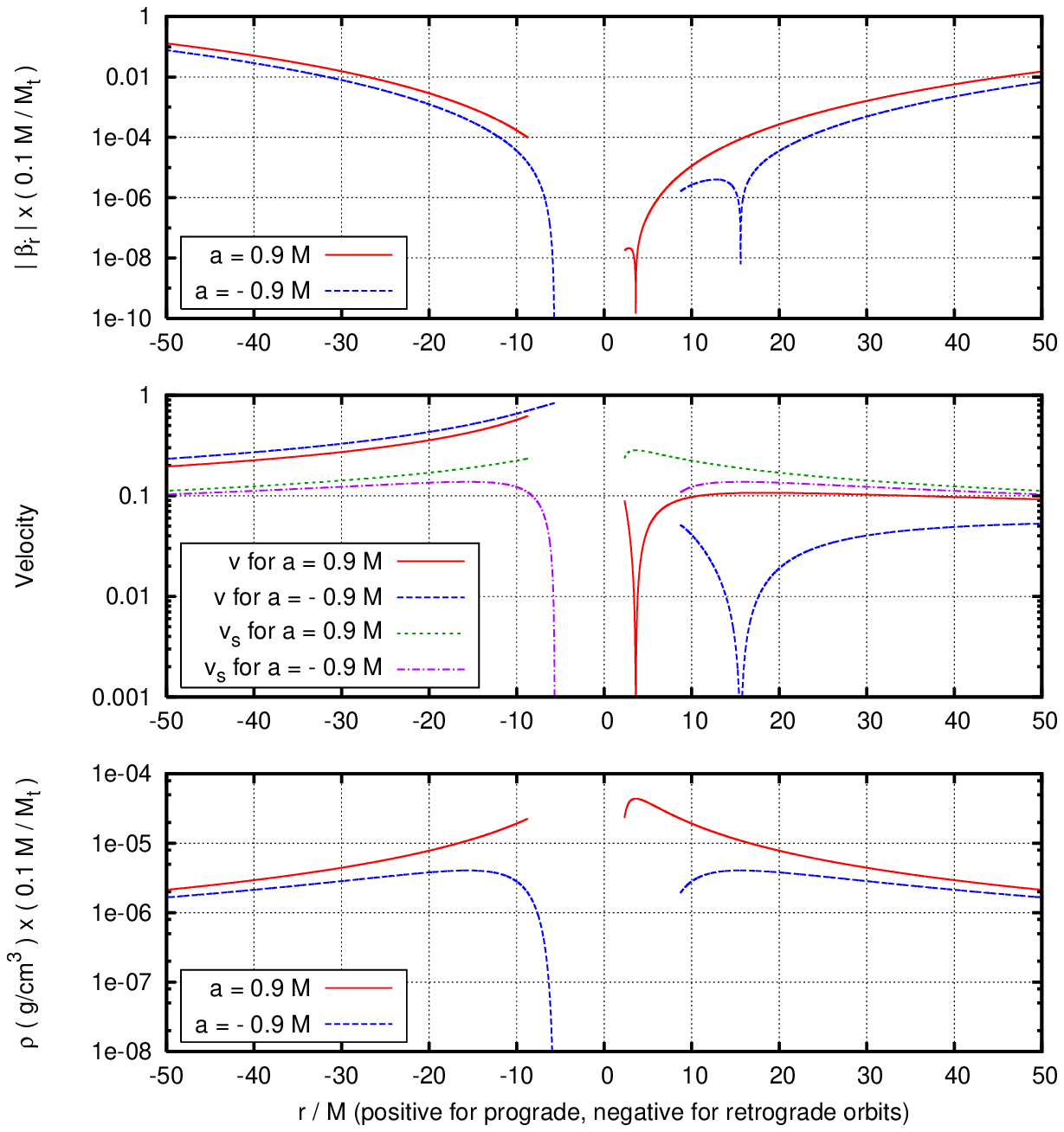}
\caption{The top plot shows the absolute value of the ratio
$\beta_{\dot{r}}\equiv(dr/dt)_{\rm hydro}/(dr/dt)_{\rm GW}$ between the rates of change of
the orbital radius due to hydrodynamic drag and radiation reaction, for circular bound stable 
orbits in the equatorial plane, as a function of the radius $r$.
For graphical reasons $r$ is considered
positive for orbits in the positive $\phi$-direction (``prograde
orbits'' \textit{i.e.,} corotating with the torus) and negative for
those in the negative $\phi$-direction (``retrograde orbits''
\textit{i.e.,} counter-rotating with respect to the torus). 
Note that all the curves of this figure (including those of the middle and bottom plots)
are terminated at the (prograde or retrograde) ISCO.
The middle plot compares the velocity $v$ of the satellite in the rest frame of
the fluid with the sound velocity $v_s$, while the bottom plot shows the
energy density of the torus.  The curves refer to the models A1 and A2
of Table \ref{table:models}, which are labeled here with the spin
parameter $a$ of the SMBH. Note that in all the plots
the vertical axis is drawn in logarithmic scale. As such, the vertical
asymptotes appearing in these plots actually correspond to a zero
value for the quantity under consideration. 
\label{fig:eq_circ_fig}}
\end{figure*}
In this section we will consider constant-$\ell$ tori around Kerr
SMBHs and compare their influence on EMRIs with that of gravitational
wave emission (\textit{i.e.,}~radiation reaction) in the adiabatic
approximation.  In particular, we will compute the rates of change
\eqref{eq:Eflux2} of the energy, angular momentum and Carter constant
due to the hydrodynamic drag, for circular equatorial orbits
(Sec. \ref{sec:eq_circ}) and for generic (inclined and eccentric)
orbits (Sec. \ref{sec:generic}).  Since $\tilde{E}$, $\tilde{L}_z$ and
$\tilde{Q}$ can be expressed analytically as functions of the orbital
parameters $p$, $e$ and $\theta_{\rm inc}$~\cite{schmidt}, it is then
easy to compute the rates of change $dp/dt$, $de/dt$ and $d\theta_{\rm
inc}/dt$ due to the hydrodynamic drag.  For the same orbits, we will
consider also the radiation reaction, for which we will use the kludge
fluxes $d\tilde{E}/dt$, $d\tilde{L}_z/dt$ and $d\tilde{Q}/dt$ of
Ref.~\cite{GGfluxes} to compute $dp/dt$, $de/dt$ and $d\theta_{\rm
inc}/dt$. Note that these kludge fluxes are a good approximation to
the fluxes computed rigorously in the adiabatic approximation with the
Teukolsky-Sasaki-Nakamura formalism~\cite{drasco_GW,TSN}. In fact,
since they are based on a Post-Newtonian expansion corrected with fits
to fluxes computed with the Teukolsky-Sasaki-Nakamura formalism for
circular orbits, these kludge fluxes are accurate within $3\%$ for
circular orbits and their accuracy is expected to be within $10-15\%$
also for generic orbits with $p\gtrsim6M$. Moreover, they are expected
to be off at most by $25-30\%$ even for smaller values of the semi-latus
rectum $p$ (\textit{cf.} Ref.~\cite{GGfluxes}, Table I).

The mass of the SMBH is fixed to $M=10^6 M_\odot$ while its
spin parameter $a$ ranges from $-0.998 M$ to $0.998 M$ (note that
$\vert a \vert=0.998 M$ is a reasonable upper limit for the spin
attainable as the result of mass accretion~\cite{thorne_spin} or
binary black-hole
mergers~\cite{rezzollaetal:2007a,rezzollaetal:2007b}), and the mass of
the satellite black hole is instead $m=1M_\odot$. The constant-$\ell$
torus is assumed to be composed of an isentropic monatomic gas
(\textit{i.e.,} a $\Gamma=5/3$ polytrope) and is considered to be
exactly filling its outermost closed equipotential surface ($\Delta
W=0$), so as to present a zero accretion rate $\dot M$ onto the
SMBH.\footnote{While realistic thick disks are generally expected to
accrete onto the SMBH, these configurations are clearly non-stationary and cannot therefore 
be reproduced within our framework. However, it is easy to show that if one cuts-off a torus solution with 
$\dot M>0$ at $r=r_{\rm cusp}$, the effect of the satellite-torus interaction will be 
enhanced with respect to the $\dot M=0$ solution having the same mass and outer radius:
the choice $\dot M=0$ is thus useful to obtain at least a lower limit for
the effect of the satellite-torus interaction on EMRIs.}
Once assumed $\Delta W=0$ and $\Gamma=5/3$, the specific angular momentum of the
torus is uniquely fixed by choosing the outer radius.  A reasonable
outer radius for a realistic accretion disk is given by $r_{\rm
out}=10^5 M$~\cite{Hure}, and this is indeed the value that we will use for most
of our analysis, although we will briefly consider also different values
for $r_{\rm out}$ in order to study the impact of this parameter on
the final results.  The polytropic constant $\kappa$ of the equation
of state is finally fixed by the requirement that $M_{\rm t}=0.1 M$.
While this could be a reasonable value for the mass of a realistic
accretion disk in AGNs~\cite{Hure}, we will see that our results scale
proportionally to $M_{\rm t}$, thus allowing one to extrapolate
them easily to the case $M_{\rm t}= M$, which
is certainly an astrophysically plausible value, but one for which our
test-fluid approach is no longer valid.

We should stress, however, that our results, when expressed in terms
of the dimensionless orbital parameters $p/M$, $e$ and $\theta_{\rm
inc}$, are approximately independent of the mass $M$ of the SMBH and
of the mass $m$ of the satellite black hole (provided that $M_{\rm
t}/M$ and $r_{\rm out}/M$ are maintained constant). Indeed, since the
ratios between the rates of change $dp/dt$, $de/dt$ and $d\theta_{\rm
inc}/dt$ due to the hydrodynamic drag and radiation reaction are of
course dimensionless, it is not restrictive to fix $M=1$, because this
simply corresponds to choosing a system of units. Note in particular
that this means that systems with different $m$ and $M$ but equal mass
ratio $m/M$ give exactly the same ratios between the rates of change
$dp/dt$, $de/dt$ and $d\theta_{\rm inc}/dt$ due to the hydrodynamic
drag and radiation reaction.  Moreover, these rates of change are
proportional to $m$ (in the case of the hydrodynamic drag this can be
seen from Eqs. \eqref{eq:bondi_rate}, \eqref{eq:tang_drag},
\eqref{eq:rad_drag}, \eqref{eq:Edot_bondi}, \eqref{eq:Ldot_bondi} and
\eqref{eq:Qdot_bondi}, while in the case of radiation reaction see for
instance Ref.~\cite{GGfluxes}), so this dependence on $m$ cancels out
when taking the ratio. The only dependence on $m$ arises from the
cutoff $r_{\min}\sim 2m (1+v^2)/v^2$ appearing in Eq.~\eqref{eq:Iphi}, but
this dependence clearly comes about only for supersonic velocities and
is a logarithmic one. As such, the results which we present in this
paper, although derived in the case of $m=1 M_\odot$ and $M=10^6
M_\odot$, are also valid for $m=0.1 M_\odot$ and $M=10^5 M_\odot$
(exactly) or for $m=1 M_\odot$ and $M=10^5 M_\odot$ (exactly for
subsonic motion, and approximately -- with an error comparable with
those affecting the fit ~\eqref{eq:Iphi} or the kludge fluxes -- for
supersonic motion).

In all of our analysis we will focus on the region close to the
SMBH ($r\lesssim50M$), which contains only a small fraction of the mass
of the torus (\textit{e.g.,} in the case of the model A1 of Table~\ref{table:models},
the mass contained in a radius $r=50M$ amounts to about $2.9\times10^{-5} M_{\rm t}$, and this fraction scales
approximately as $r_{\rm out}^{-3/2}$ when considering tori with different outer radii).  
This is the region relevant for
gravitational-wave experiments like LISA. In particular, an EMRI's
signal is expected to be detectable by LISA when its frequency (which
is twice the orbital frequency) increases above $\approx 2$ mHz (below
this frequency, in fact, there is a strong unresolvable foreground
noise due to double white-dwarf binaries in our
Galaxy~\cite{sensitivity}). This translates into a distance from the
SMBH of $r\approx 10 M$ for $M=10^6 M_\odot$, and to $r\approx
45 M$ for $M=10^5 M_\odot$. 

It should be noted, however, that the amplitude of an EMRI's signal
scales with the distance from the SMBH: for a circular orbit
of radius $r$, the Keplerian frequency is $2\pi\nu = M^{1/2}/(r^{3/2} \pm aM^{1/2}) \approx  M^{1/2}/r^{3/2}$
and the amplitude of the signal is $h\sim (m/D) (2\pi\nu M)^{2/3}\sim
(m/D)(M/r)$~\cite{peters_mathews}, where $D$ is the distance from the
observer to the source. As such, an EMRI around a $10^5 M_\odot$ SMBH
will have a gravitational-wave amplitude that at $r\sim 45 M$ is about
10 times smaller than at $r\sim 5 M$. Therefore, to see the details of
the waveforms at $r\sim 45 M$ the source must be $\sim10$ times closer
to us, which translates into a detection volume decreased by a factor
$\sim 1000$.  Nevertheless, this decrease of the detection volume may
be compensated (at least partly) by the fact that the event-rate
estimates consider only EMRIs in the strong-field region of the SMBH,
even when $M=10^5 M_\odot$~\cite{gair_event_rates}. As such, since
EMRIs in the early part of the inspiral are more numerous than those
in the strong-field region, one expects to see a number of these
events \textit{larger} than the naive estimate given by the rate
expected for strong-field EMRIs around a $10^5 M_\odot$ SMBH divided
by the detection volume decrease factor $\sim 1000$.  Of course, the
event rates could be even larger if the satellite were a
black hole with $m\sim 100 M_\odot$, because the amplitude of the
signal is proportional to $m$, but too little is presently known about
these objects to draw any sound conclusions (see for instance
Ref.~\cite{emri_rev} for a review on intermediate-mass black holes as
possible sources for LISA).

\subsection{\label{sec:eq_circ} Circular equatorial orbits}

\begin{figure*}
\includegraphics[width=14.5cm]{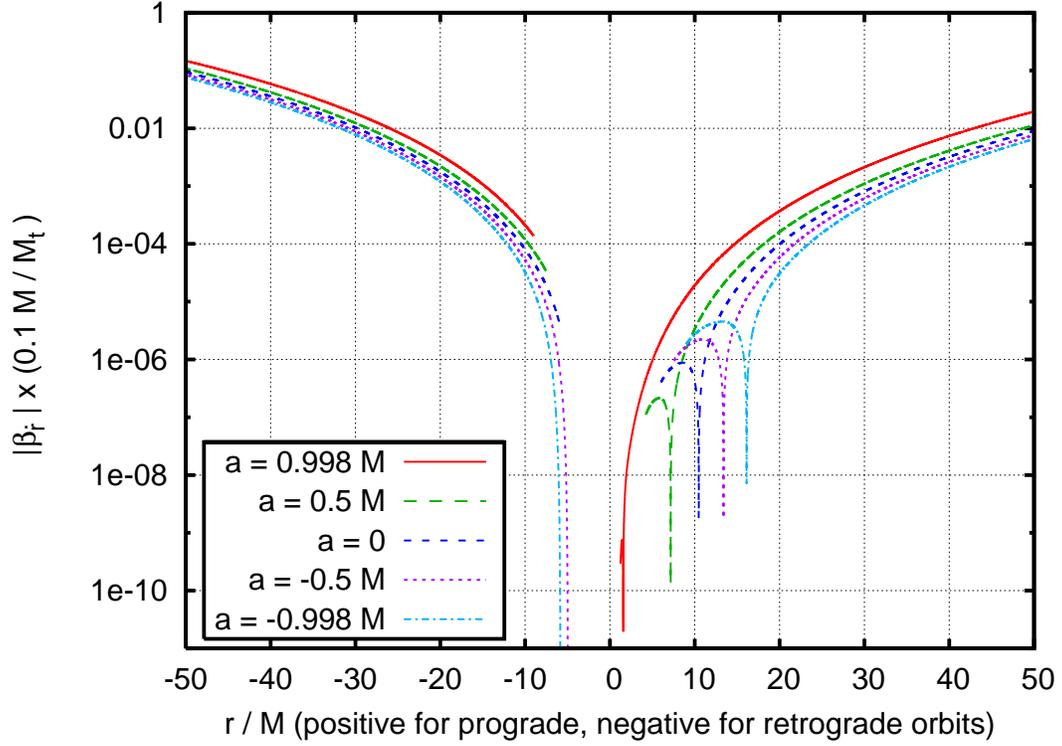}
\caption{The same as the top plot of Fig. \ref{fig:eq_circ_fig},
but for the models B1-B5 of Table \ref{table:models}, which are
labeled here with the spin parameter $a$ of the SMBH.
\label{fig:many_models}}
\end{figure*}
\begin{figure*}
\includegraphics[width=14cm]{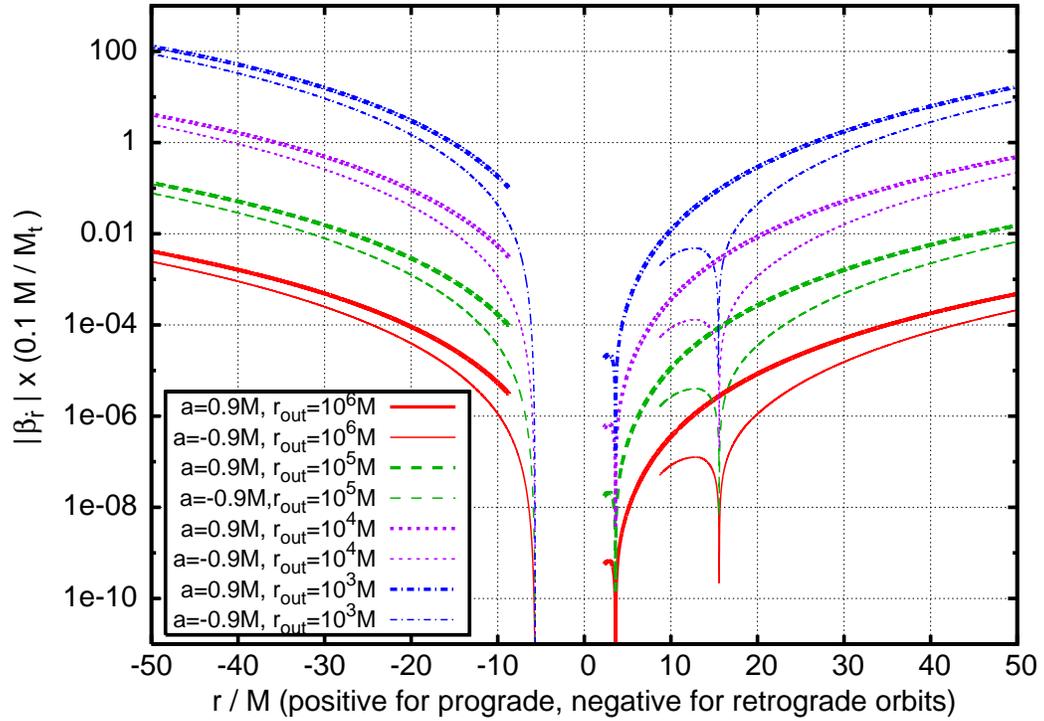}
\caption{The same as the top plot of Fig. \ref{fig:eq_circ_fig},
but for the models A1-A2 and C1-C6 of Table \ref{table:models}, which
are labeled here with the spin parameter $a$ of the SMBH and the outer radius $r_{\rm out}$ of the torus.
\label{fig:different_radii}}
\end{figure*}

The evolution of circular equatorial orbits is very simple in the
adiabatic approximation.  As mentioned in Sec. \ref{sec:interaction},
both the radiation reaction and the hydrodynamic drag maintain
circular orbits circular and, due to the symmetry of the Kerr
spacetime and of the torus with respect to the equatorial plane,
equatorial orbits will remain equatorial.  Therefore, the evolution of
circular equatorial orbits under both radiation reaction and
hydrodynamic drag can be characterized with only one quantity (the
rate of change of the radius $dr/dt$), to which the rates of change of
the energy and angular momentum,
$d\tilde{E}/dt=(d\tilde{E}/dr)\,(dr/dt)$ and
$d\tilde{L}_z/dt=(d\tilde{L}_z/dr)\,(dr/dt)$, are proportional.
($d\tilde{Q}/dt$ is instead identically zero for equatorial orbits.)
Moreover, one does not need to compute the infinite-time averages
\eqref{eq:Eflux2}, because the rates of change of $\tilde E$,
$\tilde{L}_z$ and $\tilde{Q}$ due to the hydrodynamic drag, given by
Eqs. \eqref{eq:Edot_bondi}, \eqref{eq:Ldot_bondi} and
\eqref{eq:Qdot_bondi} [or \eqref{eq:Qdot_bondi2}], are already
functions of the orbital radius alone.  [Note also that the 4-vector
$\boldsymbol \chi$ reduces to ${\boldsymbol
\chi}=-\partial_r/\sqrt{g_{rr}}$.]

The ratio between the rates of change of the orbital radius due to
hydrodynamic drag and radiation reaction is a convenient measure of
the ``efficiency'' of the hydrodynamic drag.  Defining this quantity
simply as $\beta_{\dot r} \equiv (dr/dt)_{\rm hydro}/(dr/dt)_{\rm
GW}$, we show in the top plot of Fig. \ref{fig:eq_circ_fig} the
absolute value of $\beta_{\dot r}$ as a function of the radius $r$ of
circular equatorial bound stable orbits.  The two curves refer to
models A1 and A2 of Table \ref{table:models}, and are labeled with the
spin parameter $a$ of the SMBH. Note that in order to
present all the data in a single plot, a positive $r$ refers to
orbits rotating in the positive $\phi$-direction (``prograde orbits'' \textit{i.e.,} corotating with the torus), while
a negative $r$ refers to orbits rotating in the negative $\phi$-direction (``retrograde orbits'' \textit{i.e.,}
counter-rotating with respect to the torus).  The middle plot compares
the velocity $v$ of the satellite in the rest frame of the fluid with
the sound velocity $v_s$, while the bottom plot shows instead the energy
density $\rho$ of the torus.  Note that in all the plots the vertical
axis is drawn in logarithmic scale. As such, the vertical asymptotes
appearing in Fig. \ref{fig:eq_circ_fig} (as well as in
Figs. \ref{fig:many_models} and \ref{fig:different_radii}, which are
in logarithmic scale too) actually correspond to a zero value for the
quantity under consideration.

Note that if the torus is corotating with the black hole, the radius
of the innermost stable circular orbit (ISCO) is always larger than
the inner radius of the torus, both for prograde and retrograde
orbits. For prograde orbits, this immediately follows from the
considerations of Sec. \ref{sec:torus} (since our tori have $\Delta
W=0$ and $\ell_{\rm ms}<\ell<\ell_{\rm mb}$, we have $r_{\rm
in}=r_{\rm cusp}<r_{\rm ms}$, and $r_{\rm ms}$ is exactly the radius
of the prograde ISCO), while for retrograde orbits it is sufficient to
note that the retrograde ISCO is located at a larger radius than the
prograde one.  Bearing this in mind, it is then easy to understand why
none of the quantities plotted in Fig. \ref{fig:eq_circ_fig} for model
A1 ($a=0.9 M$) goes to zero when approaching the SMBH:
although the density, the velocity of the satellite relative to the
torus, the sound velocity and $(dr/dt)_{\rm hydro}$ are exactly zero
at the inner radius of the torus, this radius is smaller than that of
the ISCO and therefore no bound stable orbits exist there.

If instead the torus is counter-rotating with respect to the black
hole (\textit{i.e.,} $a<0$), the radius of the ISCO is larger than
$r_{\rm in}$ for prograde orbits (this follows again from $r_{\rm in}=r_{\rm cusp}<r_{\rm ms}$), 
but it is not possible to conclude that the radius of the ISCO is larger than $r_{\rm in}$
also for retrograde orbits. In fact, the ISCO counter-rotating with
respect to the torus (\textit{i.e.,} the ``retrograde'' ISCO) is
corotating with the black hole and thus lies at a radius smaller than
the ``prograde'' ISCO. Indeed, for the model A2 ($a=-0.9 M$)
considered in Fig.~\ref{fig:eq_circ_fig} the retrograde ISCO is at a
radius smaller than $r_{\rm in}$. As a consequence, the density, the
sound velocity and $(dr/dt)_{\rm hydro}$ for model A2 go to zero when
the radius of the retrograde orbits decreases, being in fact zero at the
inner edge of the torus. (Of course, the velocity of the satellite
relative to the fluid does not go to zero when the radius of the
retrograde orbits decreases, because the satellite and the torus are
rotating in opposite directions.)

As it can be seen in the top plot, the ratio $\vert \beta_{\dot
r}\vert$ is larger  for the retrograde orbits than for the prograde
ones. The reason can be easily understood from the middle plot, which
shows that the retrograde motion is always supersonic.  As such, the
long-range drag, which increases significantly when passing from the
subsonic to the supersonic regime [\textit{cf.} Eq. \eqref{eq:Iphi}],
enhances the torus-satellite interaction for the retrograde
orbits. From the middle plot one can also note that relativistic
velocities ($v\gtrsim0.6$) are reached in the case of retrograde
orbits very close to the SMBH, thus further enhancing the hydrodynamic drag because of the
relativistic correction factor $\gamma^2(1+v^2)^2$ appearing in
Eq. \eqref{eq:tang_drag}. 
However, we should note that when $v$ reaches its maximum
value (\textit{i.e.,} $v\sim0.8$ for model A2) the effect of the
relativistic correction factor on the hydrodynamic drag is hindered by
the small value of the density, which goes to zero at the inner edge
of the torus (\textit{cf.} the bottom plot).

As it can be easily understood from the formulas reviewed in Sec
\ref{sec:torus}, a change in the polytropic constant $\kappa$ leaves
all the parameters of the torus unchanged, except the energy density
$\rho$, the rest mass density $\rho_0$, the pressure $p$ [all of which
scale proportionally to $\kappa^{-1/(\Gamma -1)}$] and the total mass-energy 
and rest-mass,  $M_{\rm t}$ and $M_{{\rm t} 0}$ (which scale
proportionally to $\rho$).  For this reason, the bottom plot of
Fig. \ref{fig:eq_circ_fig} scales linearly with the mass of the
torus (\textit{cf.} the label of the vertical axis). Noting, from the
formulas of Sec. \ref{sec:interaction}, that the rates of change of
the $\tilde{E}$, $\tilde{L}_z$ and $\tilde{Q}$ are proportional to the
energy density $\rho$, the same scaling applies to the top
plot. This is a very useful feature, because although the value used
for the figures of this paper -- \textit{i.e.,} $M_{\rm t}=0.1 M$ 
-- could be a plausible mass for the torus, very little is
known about these objects and larger or smaller masses may be
possible. In general, a different mass $M_{\rm t}$ could have
important effects. For instance, extrapolating to the case $M_{\rm
t}=M$, in which our test-fluid approximation is no longer valid, 
the ratio $\vert \beta_{\dot r} \vert $ would be enhanced by a
factor $10$, and for orbits counter-rotating with respect to the torus
with $r\gtrsim40 M$ the effects of hydrodynamic drag and radiation
reaction would become comparable.\footnote{Note, however, that even values of $\vert \beta_{\dot
r}\vert$ less than 1 can produce features detectable by LISA, because the 
dephasing time scales as $\sim \vert\dot r_{GW}\vert^{-1/2}$~\cite{drasco_proc}. For instance,
if $\vert\beta_{\dot r}\vert\approx 0.1$ the dephasing time between a waveform with only radiation reaction included and one
with also the effect of the hydrodynamic drag is expected to be only $\sim 3$ times
larger than the dephasing time between waveforms with and without radiation reaction included.}

It is also worth pointing out that at the center of the torus
$(dr/dt)_{\rm hydro}$ changes sign for prograde orbits, being
negative for $r>r_{\rm center}$ and positive for $r<r_{\rm
center}$. [$(dr/dt)_{\rm GW}$ is instead always negative because
gravitational waves carry a positive amount of energy away from the
source.] This change of sign corresponds, in the top plot of
Fig. \ref{fig:eq_circ_fig}, to the zero value for $\vert\beta_{\dot
r}\vert$. This behavior comes about because, although the density
reaches its maximum at the center, the motion of the fluid is exactly
Keplerian (geodesic) there, and the relative velocity of the satellite
is therefore exactly zero (\textit{cf.} the middle plot). This means
in particular that $\tilde{E}=-u^{\rm sat}_t=-u_t^{\rm fluid}=U$,
which together with Eq. \eqref{eq:Edot_bondi} and $\chi_t=0$ gives
$\dot{\tilde{E}}=0$ and therefore $\dot{r}=0$ for prograde orbits at
$r=r_{\rm center}$. Moreover, if $r>r_{\rm center}$ the specific
angular momentum of the satellite is larger than that of the torus 
(\textit{cf.} for instance Ref.~\cite{constant_l_disks1}, Fig. 5), and therefore the
satellite is slowed down by the interaction with the fluid
(\textit{i.e.}, $\dot{r}<0$). On the other hand, if $r<r_{\rm
center}$ the specific angular momentum of the satellite is smaller
than that of the torus, and the satellite is speeded-up (\textit{i.e.},
$\dot{r}>0$).

%--->

Fig. \ref{fig:many_models} shows the absolute value of $\beta_{\dot
r}$ as a function of the radius $r$ of circular equatorial bound
stable orbits for the models B1-B5 of Table \ref{table:models}, whose
spin parameter $a$ goes from $-0.998 M$ to $0.998 M$.  As it can be
seen, the situation is qualitatively very similar to the one presented
in the top plot of Fig. \ref{fig:eq_circ_fig}. In particular, the
effect of the hydrodynamic drag can be comparable to that of radiation
reaction, but only if we extrapolate to $M_{\rm t}=M$ and, even in
that case, only for orbits counter-rotating with respect to the torus
and with $r\gtrsim40 M$.

We can also note that the effect of the spin $a$ on the results is
negligible, except for the prograde orbits between the center and the
ISCO, for which $\vert\beta_{\dot r}\vert$ decreases as $a$ increases.
The reason for this can be easily understood by 
considering a satellite moving on a prograde circular equatorial orbit between the center and
the inner edge of the torus, and by recalling that the
difference between $\ell_{\rm mb}$ and $\ell_{\rm ms}$ represents an
upper limit for the deviation of the specific angular momentum of the
satellite away from that of the torus (see
Ref.~\cite{constant_l_disks1}, Fig. 5). Because this deviation
regulates the exchange of angular momentum between the torus and the
satellite [\textit{cf.} Eq.~\eqref{eq:Ldot_bondi}, where $\chi_\phi=0$ for circular orbits]
and thus the rate of change of the orbital radius, 
$\beta_{\dot r}$ must go to zero if $\ell_{\rm
mb}-\ell_{\rm ms}$ goes to zero. Since it is easy to verify
that $\ell_{\rm mb}-\ell_{\rm ms} \to 0$ as $a\to M$ [\textit{cf.}
Eq. \eqref{eq:l_ms_l_mb}], it is natural to find that
$\vert\beta_{\dot r}\vert$ decreases as $a$ increases.

Finally, in Fig.~\ref{fig:different_radii} we plot again $\vert
\beta_{\dot r}\vert$, but for models A1-A2 and C1-C6 of Table
\ref{table:models}, in which we have considered different values for
the outer radius $r_{\rm out}$ of the torus, ranging from $10^3 M$ to
$10^6 M$. The reason for this is that although $r_{\rm out}\sim 10^5
M$ is a plausible value for the outer radius, little is
known about the size of astrophysical accretion disks and larger
or smaller outer radii may also be possible. As it can be seen from
Fig. \ref{fig:different_radii}, a different outer radius will have
significant effects for prograde orbits with $r\gtrsim20 M$ and
retrograde orbits with $r\gtrsim10 M$, for which the effect of the hydrodynamic
drag can become comparable to that of radiation reaction.
In general, $\vert\beta_{\dot r}\vert$ progressively increases as the outer radius is decreased. This is
rather simple to explain: decreasing $r_{\rm out}$ while keeping
$M_{\rm t}$ constant amounts to increasing the average rest-mass
density and hence the hydrodynamic drag.  When considered from this
point of view, the uncertainty on the value of $r_{\rm out}$ has an effect opposite
to the the uncertainty about the mass of the torus: a decrease of $r_{\rm out}$ (or an increase of $M_{\rm t}$) 
induces an increase of $\vert\beta_{\dot r}\vert$. For circular orbits, this
overall uncertainty can be easily modelled in terms of a simple
scaling of the type
\begin{equation}
\label{scaling}
\beta_{\dot r}(r_{\rm out}) \approx \beta_{\dot r}\vert_{5} 
\left(\frac{10^5\,M}{r_{\rm out}}\right)^{3/2}\,,
\end{equation}
where $\beta_{\dot r}\vert_{5}$ is the efficiency for
  $r_{\rm out} = 10^5\,M$. Note that the scaling power is not $3$ as
  one may naively expect. This is because $\beta_{\dot r}$ is most
  sensitive to the changes of the rest-mass density in the inner part
  of the torus and this does not scale simply as $r_{\rm out}^{-3}$.

\begin{figure*}
\includegraphics[angle=-90, width=17cm]{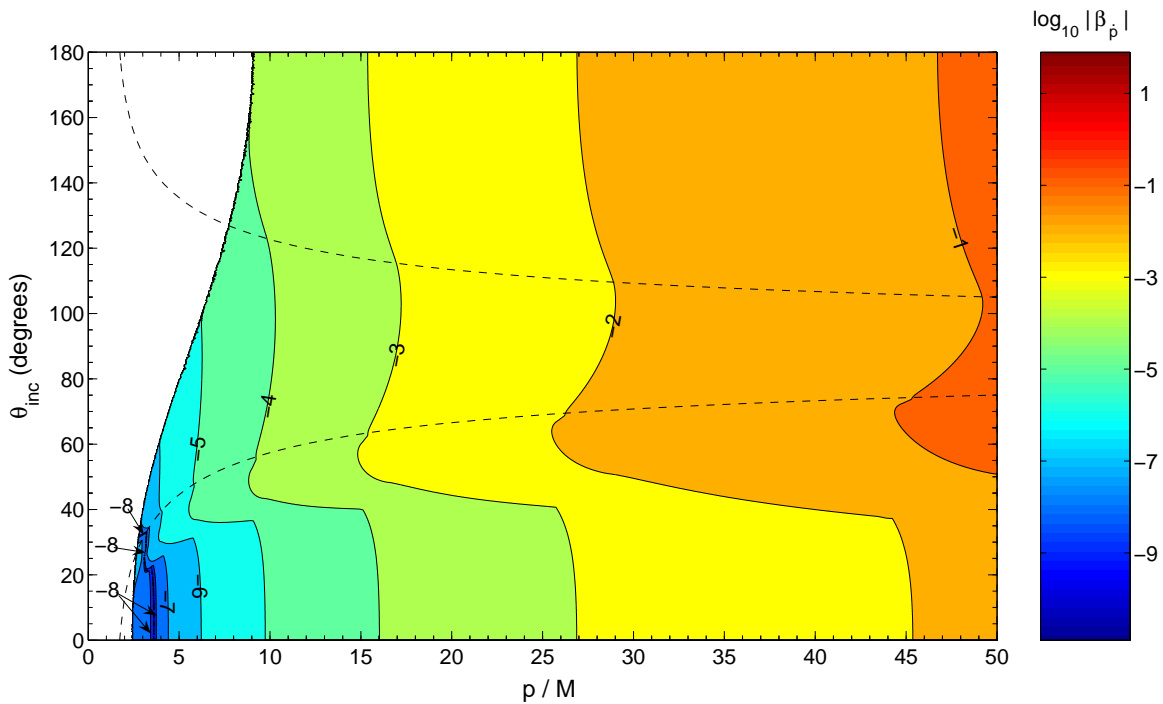}
\caption{$\log_{10}\vert\beta_{\dot p}\vert$ for inclined orbits
with $e=0.1$ as a function of the semi-latus rectum $p$ and of the
inclination angle $\theta_{\rm inc}$.  The figure refers to the model
A1 of Table \ref{table:models}, but a larger (smaller) mass $M_{\rm
t}$ for the torus would simply increase (decrease) the absolute ratio
$\vert\beta_{\dot p}\vert$ by a factor $M_{\rm t}/(0.1 M)$. The dashed
line marks the edge of the torus.
\label{fig:pdot}}
\end{figure*}

\subsection{\label{sec:generic} Generic orbits}

We will now extend the analysis of Sec. \ref{sec:eq_circ} to bound
stable generic (inclined and eccentric) orbits.  Although such an
extension is in principle straightforward using the formulas
introduced in Sec. \ref{sec:adiabatic}, the space of parameters and
results which one needs to examine greatly enlarges. Not only are generic
orbits characterized by the three parameters $p$, $e$,
$\theta_{\rm inc}$ defined by Eq. \eqref{eq:pei_defs}, but one also
needs to consider three quantities describing the evolution of each
single orbit in the parameter space \textit{i.e.,} the rates of change
$dp/dt$, $de/dt$ and $d\theta_{\rm inc}/dt$.

To simplify our analysis, we will focus mainly on model A1 of
Table~\ref{table:models}, which could be a representative example of an
astrophysical torus in an AGN, and then examine how the rates of change
$dp/dt$, $de/dt$ and $d\theta_{\rm inc}/dt$ due to the hydrodynamic drag compare to
those due to radiation reaction throughout the
space of parameters $(p,e,\theta_{\rm inc})$. 
The considerations that we will draw for model A1 can, however,
be extended simply to the cases of different masses and radii for the
torus. As in the case of circular orbits, in fact, a larger (smaller)
mass $M_{\rm t}$ for the torus when $r_{\rm out}$ is held constant
would simply increase (decrease) the rates $dp/dt$, $de/dt$ and
$d\theta_{\rm inc}/dt$ due to the hydrodynamic drag by a factor
$M_{\rm t}/(0.1 M)$. This scaling is \textit{exact} (as long as the torus is not self-gravitating) and comes about because the 
rates of change of $\tilde{E}$, $\tilde{L}_z$ and $\tilde{Q}$ (and consequently those
of $p$, $e$ and $\theta_{\rm inc}$) are proportional to the energy density $\rho\propto M_{\rm t}$. 
Similarly, variations of $r_{\rm out}$ will
result in an effect which is similar to the one discussed for
Fig. \ref{fig:different_radii} in the case of circular orbits
[\textit{cf.}, eq.~\eqref{scaling}], as we will see at the end of this section.

All of the results presented in this Section have been computed by
integrating numerically Eq. \eqref{eq:Eflux2} using an iterated
Romberg method~\cite{nr}, with a typical accuracy, depending on the
parameters of the orbit under consideration, of $10^{-7}$--$10^{-4}$
and never worse than $4\times10^{-3}$.~\footnote{Note that the
accuracy of the numerical integration is certainly adequate, because
it is considerably better than the errors affecting the fit \eqref{eq:Iphi}
as well as those affecting the kludge fluxes that we use to study the
effects of radiation reaction.}

\begin{figure*}[t]
\includegraphics[angle=-90, width=17cm]{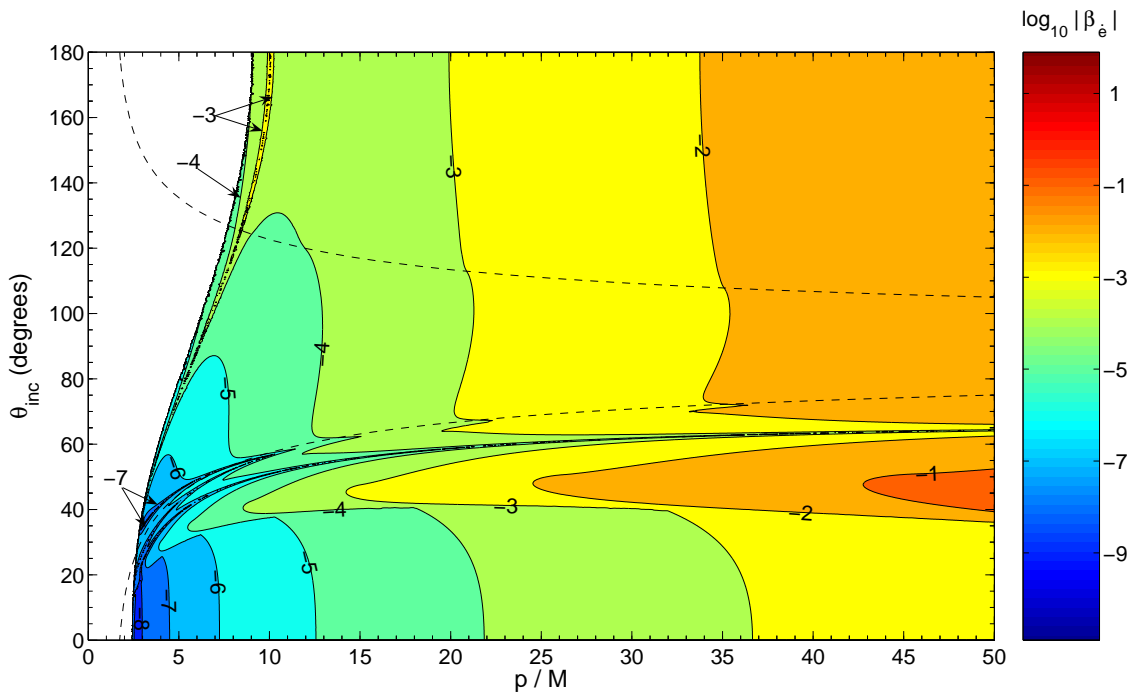}
\caption{The same as in Fig. \ref{fig:pdot}, but for
$\log_{10}\vert\beta_{\dot e}\vert$.
\label{fig:edot}}
\end{figure*}
\begin{figure*}
\includegraphics[angle=-90, width=17cm]{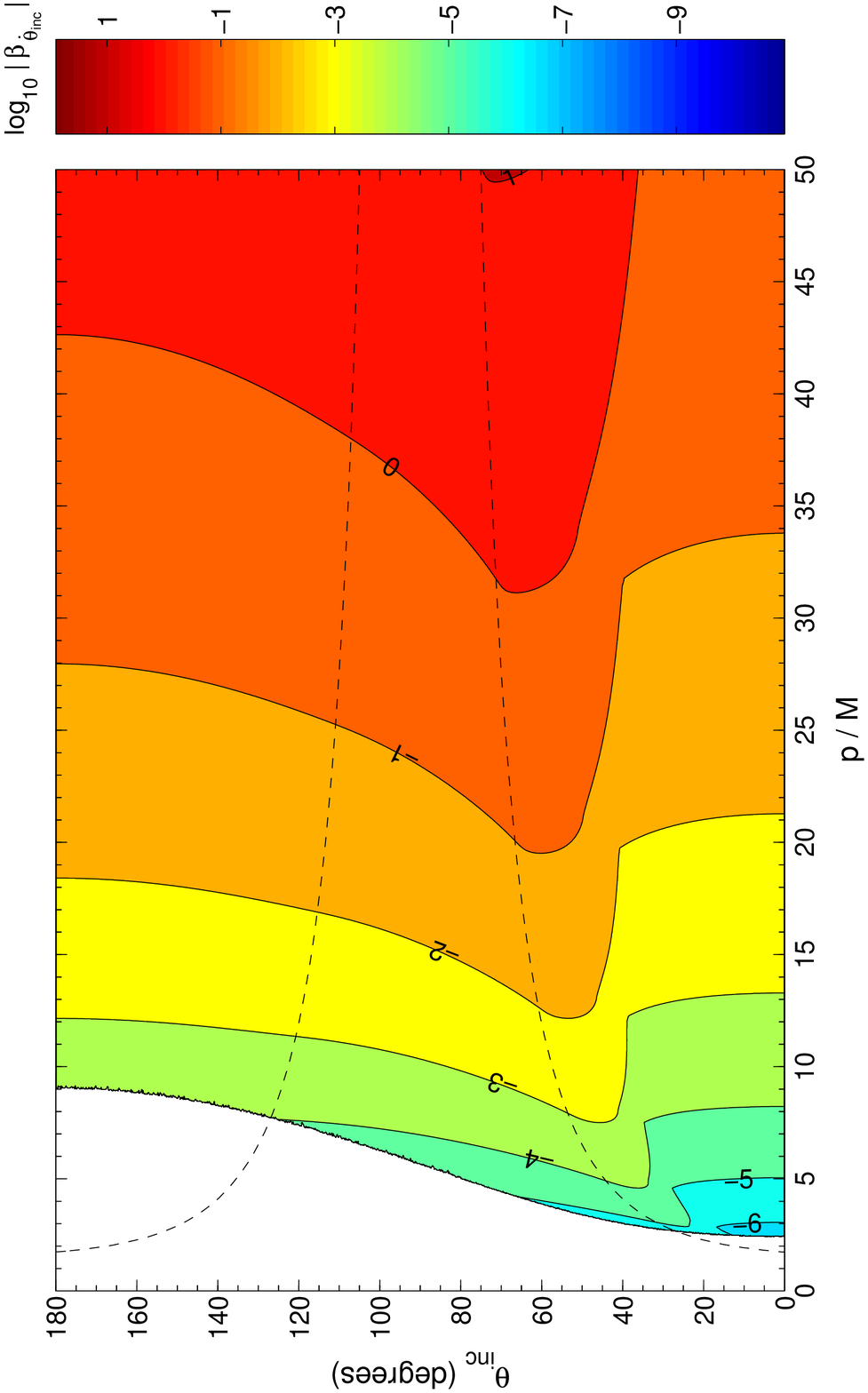}
\caption{The same as in Fig. \ref{fig:pdot}, but for
  $\log_{10}\vert\beta_{\dot \theta_{\rm inc}}\vert$.
\label{fig:cosdot}}
\end{figure*}

We start by analyzing in detail orbits with eccentricity $e=0.1$, and
we will then study the effect of a different eccentricity on the
results. In Figs. \ref{fig:pdot}, \ref{fig:edot}, \ref{fig:cosdot} the
color-code and the contour levels show the base-10 logarithm of the
``efficiencies'' of $\dot p$, $\dot e$ and $\dot{\theta}_{\rm inc}$,
\textit{i.e.}, $\beta_{\dot p} \equiv \vert(dp/dt)_{\rm
hydro}/(dp/dt)_{\rm GW}\vert$, $\beta_{\dot e} \equiv
\vert(de/dt)_{\rm hydro}/(de/dt)_{\rm GW}\vert$ and $\beta_{\dot
\theta_{\rm inc}} \equiv \vert(d\theta_{\rm inc}/dt)_{\rm
hydro}/(d\theta_{\rm inc}/dt)_{\rm GW}\vert$, as functions of
$p$ and $\theta_{\rm inc}$.  The dashed line marks the ``edge of the
torus'' \textit{i.e.,} the location in the $(p,\theta_{\rm
inc})$-plane of the orbits having $\theta_{\rm inc}=\theta_{\rm
inc,\,t}(p)$, where $\theta_{\rm inc,\,t}(r)$ is the function giving
the angle between the surface of the torus and the equatorial plane in
terms of the radius $r$. The blank part on the left portion of these
figures refers to the region where no bound stable orbits exist, and we
will refer to the line marking the boundary of this region
as the \textit{separatrix} \cite{zoom_whirl} 

Each figure has been obtained by computing the quantity under
consideration using Eq. \eqref{eq:Eflux2} for $\sim 5\times10^5$
orbits irregularly distributed in the $(p,\theta_{\rm inc})$-plane,
and then linearly interpolating on a grid of $1500\times1500$ nodes
using a Delaunay triangle-based method. The gridded data obtained in
this way has been used to draw the contour levels.  Not surprisingly,
Figs. \ref{fig:pdot}, \ref{fig:edot}, \ref{fig:cosdot} show somewhat
the same trend as the results presented in Sec. \ref{sec:eq_circ} for
circular equatorial orbits, with the effect of the torus becoming
comparable to that of the radiation-reaction far away from the black
hole and becoming instead negligible in the strong-field region of the
black hole. However, these figures present also a variety of features
that we will now analyze in detail.

Fig. \ref{fig:pdot}, for instance, shows $\vert \beta_{\dot p}\vert$
and indicates that the effect of the hydrodynamic drag is larger for orbits with
high inclination, for any given semi-latus rectum $p$. This is simply due
to the fact that orbits with $\theta_{\rm inc}> 90$ degrees are
retrograde with respect to the torus, and the velocity of the
satellite relative to the fluid can easily become supersonic. Indeed
this effect is visible also in the figures of
Sec. \ref{sec:eq_circ}. (We recall that in those figures the
retrograde orbits are mapped to negative values of the orbital radius
$r$.)  The transition between the subsonic and the supersonic regime
is marked by the sharp bend of the contour levels of
Fig. \ref{fig:pdot} at $\theta_{\rm inc}\sim 40$ degrees. This bend
corresponds to the passage from orbits which are always subsonic (the
orbits with $\theta_{\rm inc}$ smaller than the inclination angle at
which the bend is located) to orbits which are supersonic at least for
a part of their trajectory (the orbits with $\theta_{\rm inc}$ larger
than the inclination angle at which the bend is located). Another
small dip is hardly noticeable in the contour levels at inclination
angles $\theta_{\rm inc}$ just smaller than the edge of the torus (and
smaller than $90$ degrees, corresponding therefore to prograde
orbits); this feature corresponds to the transition from orbits which
are partly subsonic and partly supersonic (``below'' the dip), to
orbits which keep always supersonic (``above'' the dip).\footnote{We
note that in order to better understand the fine features in the
contour levels, we have built an auxiliary code computing the
quantities $(dp/dt)_{\rm hydro}$, $(de/dt)_{\rm hydro}$ and
$(d\cos\theta_{\rm inc}/dt)_{\rm hydro}$ by direct integration of
Eqs. \eqref{eq:Edot_bondi}, \eqref{eq:Ldot_bondi},
\eqref{eq:Qdot_bondi} and \eqref{eq:Qdot_bondi2} along numerically
solved geodesics, averaging over a reasonably large number
of revolutions  ($\sim30$) for each geodesic. This has not only validated the
results which have been used to build the figures and which have been
obtained using Eqs. \eqref{eq:Eflux2}, but has also allowed us to
examine in detail the behavior of the geodesics in the various regions
of the parameter space $(p,e,\theta_{\rm inc})$, thus helping to
interpret the complicated features of the figures shown in this
paper.}  From Fig. \ref{fig:pdot} one can also note that  $\vert \beta_{\dot p}\vert$ 
becomes lower than $10^{-8}$ in a narrow ``strip'' at $p/M<5$. Indeed, $(dp/dt)_{\rm
hydro}$ changes sign inside this ``strip'', being positive inside the
region between the ``strip'' and the separatrix and negative outside,
while $(dp/dt)_{GW}$ is always negative.  This behavior generalizes
that of circular equatorial orbits, for which $(dr/dt)_{\rm hydro}$
changes sign at the center of the torus (\textit{cf.}
Sec. \ref{sec:eq_circ}).  Also in this case, however, the very small
values of $\vert \beta_{\dot p}\vert$ cannot produce an observable imprint
on the waveforms.

In a similar way, Fig. \ref{fig:edot} shows the behavior of $\vert
\beta_{\dot e} \vert$. As it can be seen, the influence of the torus
is again larger at high inclinations than at low ones, for any fixed
semi-latus rectum. Also in this case, this happens because the orbits
counter-rotating with respect to the torus can easily become
supersonic. As in Fig. \ref{fig:pdot}, we can note the presence of a
sharp bend in the contour levels at $\theta_{\rm inc}\sim 40$ degrees,
due to the transition from orbits always subsonic to orbits partly
supersonic, and a dip in the contour levels near the edge of the torus
(at inclinations $\theta_{\rm inc}< 90$ degrees), more pronounced than
in Fig. \ref{fig:pdot} but again due to the transition from orbits
only partly subsonic to orbits always supersonic. Moreover, one can
note the presence of three ``valleys'' where the efficiency
$\vert\beta_{\dot e}\vert$ becomes very small.  One (``valley 1'')
starts at $\theta_{\rm inc}\approx 15$ degrees, very close to the
separatrix, and extends as far as the right edge of the figure
($p/M=50$, $\theta_{\rm inc}\approx65$ degrees) and beyond.  A second
valley (``valley 2'') starts at the same point as valley 1, but
extends only until $p/M\approx 12$ and $\theta_{\rm inc}\approx60$
degrees, where it terminates together with a third valley (``valley
3'') starting on the separatrix at $\theta_{\rm inc}\approx30$
degrees.  Across these valleys, the quantity $(de/dt)_{\rm hydro}$
becomes zero and changes sign, being negative under valley 1 and in
the region between the separatrix and valleys 2 and 3, and positive in
the rest of the $(p,\theta_{\rm inc})$-plane. Conversely, the rate of
change of the eccentricity due to radiation reaction is always
negative, with the exception of orbits very close to the
separatrix~\cite{GGfluxes}; this is apparent also in
Fig. \ref{fig:edot}, where the narrow ``strip'' corresponding to a
ratio $\vert\beta_{\dot e}\vert \gtrsim 10^{-3}$ and running close and
almost parallel to the separatrix is due to a change in sign of
$(de/dt)_{\rm GW}$. Despite this markedly different behavior of
$(de/dt)_{\rm hydro}$ and $(de/dt)_{\rm GW}$, Fig. \ref{fig:edot}
shows that the effect of the hydrodynamic drag is always much smaller
than radiation reaction unless the semi-latus rectum of the orbit is
increased to $p/M \gtrsim 50$, or the mass of the torus is increased
at least by a factor 10 thus extrapolating 
to $M_{\rm t}=M$\footnote{The test-fluid approximation of course breaks down in this limit.}. However, while a larger
semi-latus rectum increases the efficiency $\beta_{\dot e}$, it also
reduces the frequency and amplitude of the gravitational-wave signal,
moving it to a region of low sensitivity for LISA.

\begin{figure*}[t]
\includegraphics[angle=-90, width=17cm]{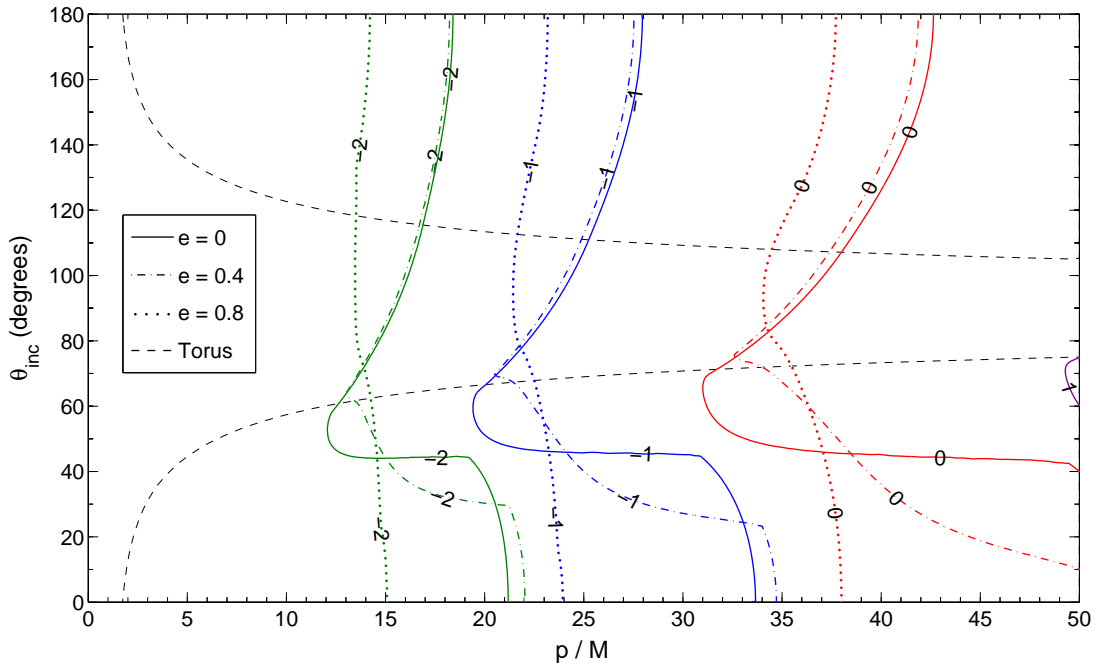}
\caption{$\log_{10}\vert\beta_{\dot \theta_{\rm inc}}\vert$ shown as a
 function of the semi-latus rectum $p$ and of the inclination angle
 $\theta_{\rm inc}$ for inclined orbits with selected values of the
 eccentricity, \textit{i.e.}, $e=0$ (solid line), $e=0.4$ (dot-dashed
 line), $e=0.8$ (dotted line). The figure refers to the model A1 of
 Table \ref{table:models}, but a larger (smaller) mass $M_{\rm t}$ for
 the torus would simply increase (decrease) $\vert\beta_{\dot
 \theta_{\rm inc}}\vert$ by a factor $M_{\rm t}/(0.1 M)$. The dashed
 line marks the egde of the torus. 
\label{fig:many_ecc_cosdot}}
\end{figure*}

\begin{figure*}[t]
\includegraphics[angle=-90, width=15.5cm]{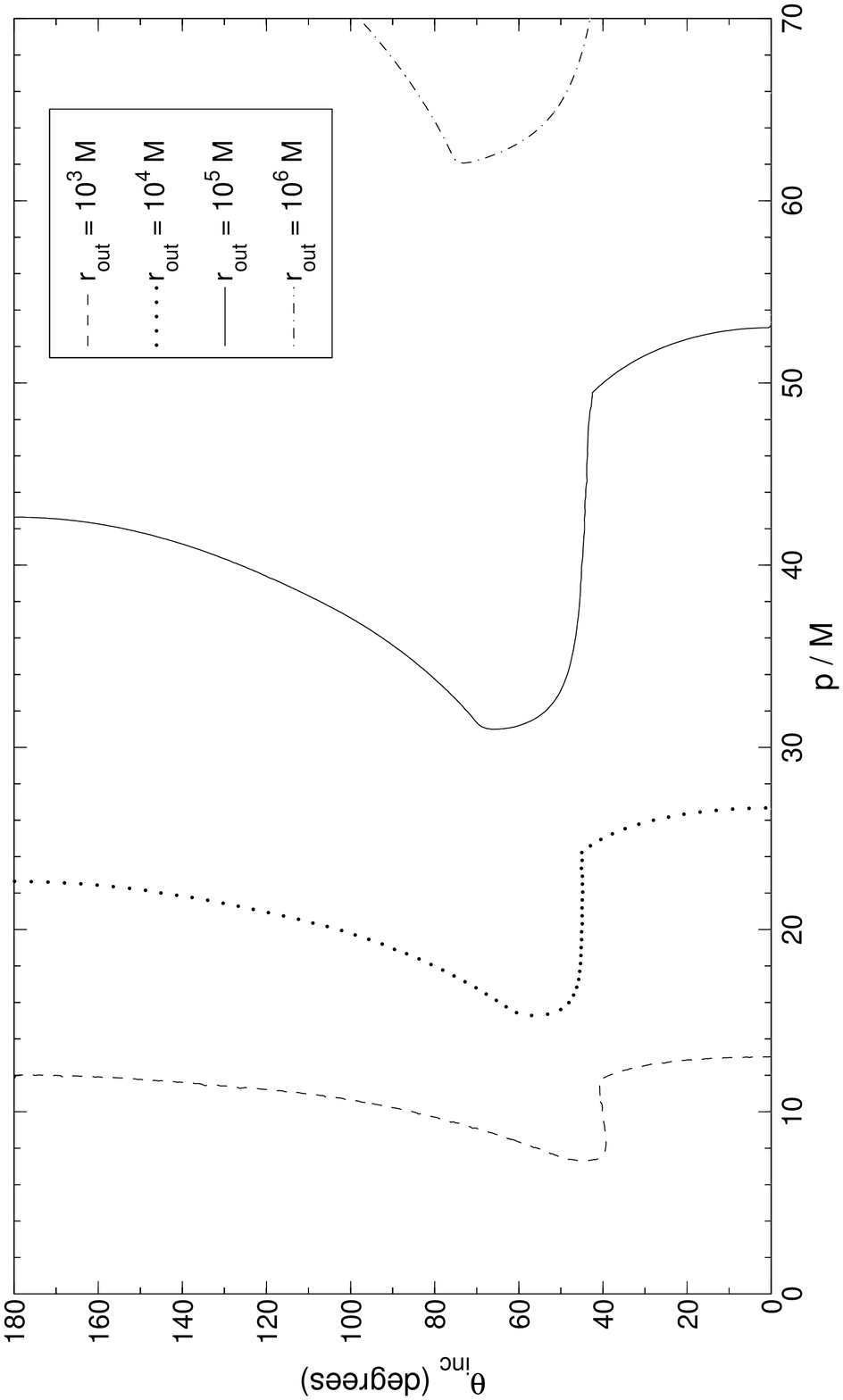}
\caption{Location in the $(p,\theta_{\rm inc})$-plane of the
\textit{circular} orbits for which $\vert\beta_{\dot \theta_{\rm
inc}}\vert = 1$. Different curves refer to different values of the
outer radius and thus to models A1, C1, C3 and C5 of Table
\ref{table:models}. Note that a different mass $M_{\rm t}$ for the
torus would simply make the curves of this figure to correspond to
$\vert\beta_{\dot \theta_{\rm inc}}\vert = M_{\rm t}/(0.1 M)$.
 \label{fig:angle_diff_radii}}
\end{figure*}

The effect of the hydrodynamic drag is somewhat stronger when
considered in terms of the efficiency $\vert\beta_{\dot \theta_{\rm
inc}}\vert$, as it is shown in Fig. \ref{fig:cosdot}.  While the
qualitative behavior is similar to the one discussed for the two
preceding figures, it should be noted that the effect of the torus is
comparable to that of radiation reaction already for $p/M \approx 32$
if $\theta_{\rm inc} \approx 60$ degrees, and the two effects remain
comparable down to $p/M\approx20$ if the mass of the torus is
increased by a factor 10 thus extrapolating to $M_{\rm t}=M$. Moreover, while radiation reaction produces
an increase in the inclination $\theta_{\rm inc}$ irrespective of the
orbital parameters~\cite{NHSO}, $(d\theta_{\rm inc}/dt)_{\rm hydro}$
is always negative and thus a measurement of the evolution of the
inclination angle $\theta_{\rm inc}$ in the early stages of an EMRI could
give important information on the presence of a torus. 
If such a presence were to be detected, it would not prevent 
high-precision tests of the Kerr nature of the SMBH being performed in the strong-field region,
where the hydrodynamic drag becomes negligible.

We should also note that the decrease of $\theta_{\rm inc}$ due to the
hydrodynamic drag is not surprising for orbits with $\theta_{\rm inc}>
90$ degrees (\textit{i.e.,} orbits rotating in the opposite
$\phi$-direction with respect to the fluid), because the hydrodynamic
drag clearly induces the orbits to rotate in the same $\phi$-direction
as the torus. For orbits with $\theta_{\rm inc}< 90$ degrees, instead,
the decrease of $\theta_{\rm inc}$ comes directly from
Eq. \eqref{eq:u_theta} (we recall that the effect of the radial drag
averages out when adopting the adiabatic approximation), thus
following directly from the axis- and plane-symmetry of the system and
being independent of the use of constant specific angular momentum
tori such as the ones considered in this paper. Indeed, since the fluid of
the torus does not move in the $\theta$-direction,
Eq. \eqref{eq:u_theta} states that accretion conserves the momentum of
the satellite in the $\theta$-direction, but it also increases its
mass, thus reducing the velocity in the $\theta$-direction. In
addition, the dynamical friction will contribute to this reduction by
damping further the oscillations around the equatorial plane.

To illustrate how the above results depend on the eccentricity, we
show in Fig.~\ref{fig:many_ecc_cosdot} the efficiency $\vert
\beta_{\dot \theta_{\rm inc}}\vert$ for the model A1 of Table
\ref{table:models}, but for different values of the eccentricity
\textit{i.e.,} $e=0$, $0.4$, and $0.8$. (Equivalent figures could be
made also for $\vert \beta_{\dot p} \vert$ and $\vert \beta_{\dot e}
\vert$, but we omit them here because they would be qualitatively similar to
Fig.~\ref{fig:many_ecc_cosdot}). For each value of the eccentricity,
we have computed $\vert \beta_{\dot \theta_{\rm inc}}\vert$ for $\sim
4\times10^4$ orbits, and using the same technique employed for
Figs. \ref{fig:pdot}, \ref{fig:edot}, \ref{fig:cosdot} we have drawn
the contours corresponding to values of $0.01$, $0.1$, $1$ and
$10$. Also in this case, a larger (smaller) mass $M_{\rm t}$ for the
torus would simply increase (decrease) these absolute ratios by a
factor $M_{\rm t}/(0.1 M)$.

Clearly, many of the features in this plot have been discussed also
for the previous figures. 
For instance, the contour levels present sharp bends at low inclinations 
(\textit{i.e.}, $\theta_{\rm inc}\approx10-40$ degrees) 
for $e=0$ and $e=0.4$, due the transition from subsonic
to partly supersonic orbits, whereas the transition from partly supersonic to fully supersonic
orbits causes the appearance of a pronounced ``kink'' in the $e=0.4$ contour levels, for inclinations
$\theta_{\rm inc}< 90$ degrees just above the edge of the torus.
The $e=0.8$ contour levels, on the other hand, are rather
smooth and less affected by the complex changes of regimes as the
satellite interacts with the torus.  Most
importantly, however, Fig.~\ref{fig:many_ecc_cosdot} suggests that the
conclusions drawn when discussing Fig. \ref{fig:cosdot} for orbits
with $e=0.1$ are not altered significantly by a change in the
eccentricity. Indeed, even for large eccentricities the influence of
the torus on the evolution of $\theta_{\rm inc}$ can be comparable to
that of radiation reaction for $p/M$ as small as $35-38$, while the
two effects are still equal at $p/M\approx23-24$ if the mass of the
torus is increased by a factor $10$ thus extrapolating to $M_{\rm t}=M$. As a result, a
measurement of the evolution of the inclination angle $\theta_{\rm
inc}$ even for generic eccentric orbits could give important
information on the presence of a torus around the SMBH.

This conclusion is finally confirmed by
Fig.~\ref{fig:angle_diff_radii}, in which we show how
$\vert\beta_{\dot \theta_{\rm inc}} \vert$ changes if one considers
different values of the outer radius of the torus while keeping its
mass fixed.  More specifically, Fig. \ref{fig:angle_diff_radii} shows
the location in the $(p,\theta_{\rm inc})$-plane of the \textit{circular}
orbits for which $\vert\beta_{\dot \theta_{\rm inc}}\vert
=1$. Different curves refer to different values of the outer radius,
and in particular to the models A1, C1, C3 and C5 of Table \ref{table:models}.
As it is probably obvious by now, a different mass $M_{\rm t}$ for the
torus would simply make the curves of this figure correspond to
$\vert\beta_{\dot \theta_{\rm inc}}\vert = M_{\rm t}/(0.1 M)$. As
expected from the results of Sec. \ref{sec:eq_circ}, a modest
variation of the outer radius can easily cause the decrease of the
inclination angle due to the hydrodynamic drag to be dominant over the
increase due to radiation reaction for orbits with $p/M\sim20$ or
smaller.

\section{\label{sec:conclusion}Conclusions}

SMBHs are expected to be surrounded by matter, either in the form of
stellar disks, as in the case of ``normal'' galactic centers, or in the
form of accretion disks of gas and dust, as in the case of AGNs. In
order to assess under what conditions and to what extent the
interaction with matter could modify the gravitational-wave signal
from EMRIs in AGNs, we have studied EMRIs in spacetimes containing a
SMBH surrounded by a non self-gravitating torus. For simplicity, and
in order to handle the equilibrium solution analytically, we have
considered a torus with a constant distribution of specific angular
momentum, using as reference dimensions and masses those for the
accretion disks expected in AGNs, but bearing in mind that these also
come with rather large uncertainties. We have extrapolated our results
also to cases in which the mass of the torus is comparable with that of the SMBH, although we stress
that in this limit our test-fluid approximation for the torus is no longer valid.

Overall, we have found that the effect of the hydrodynamic drag
exerted by the torus on the satellite black hole can have important
effects sufficiently far from the central object, and that these
effects are qualitatively different from those of radiation
reaction. In particular, if the torus is corotating with the SMBH, the hydrodynamic drag always
\textit{decreases} the inclination of the orbits with respect to the
equatorial plane (\textit{i.e.,} orbits evolve towards the
equatorial prograde configuration), whereas radiation reaction always
\textit{increases} the inclination (\textit{i.e.,} orbits evolve
towards the equatorial retrograde configuration).  In the case of a
system composed of a SMBH with mass $M=10^6 M_\odot$ and a corotating
torus with mass $M_{\rm t}\lesssim M$, the effect of the torus will be marginally
observable by LISA only if the radius of the torus is as small as
$r_{\rm out}\approx 10^3-10^4 M$. However, if the SMBH has a
lower mass, EMRIs will be detectable by LISA at larger distances from
the SMBH, and the effects of a torus will be more evident.  For
instance, for a SMBH with $M=10^5 M_\odot$ and a corotating torus with outer
radius $r_{\rm out}=10^5 M$ and mass $M_{\rm t}=0.1 M$ ($M_{\rm
t}=M$), the inclination with respect to the equatorial plane will
decrease, due to the hydrodynamic drag, for orbits with semi-latus rectum
$p\gtrsim35M$ ($p\gtrsim25 M$), while the EMRI signal will start being
detectable by LISA already at a distance of $\approx 50 M$ from the
SMBH.  Note, however, that unless one considers as the satellite an
intermediate-mass black hole with $m\sim 100 M_\odot$ around a $10^5
M_\odot$ SMBH (a configuration which may be possible but about which too
little is yet known), considering EMRIs at such large distances from
the SMBH has the obvious drawback that the amplitude of the
gravitational-wave signal will be proportionally smaller. This will
considerably reduce the detection volume, although the decrease in the
event rate could be mitigated by the fact that weak-field EMRIs are
probably more numerous than strong-field EMRIs, which are the ones
accounted for in standard calculations of event rates.

In general, we expect measurements of the evolution of the inclination
angle in the early phases of EMRIs to be a potential source of
important information about the presence of thick tori which could not
be detected by other techniques. Moreover, because for any
astrophysically plausible torus configurations the effect of the
hydrodynamic drag becomes rapidly negligible in the very strong-field
region of the SMBH (\textit{i.e.,} $p\lesssim5M$), 
the presence of a torus would not prevent high-precision tests of
the Kerr nature of the SMBH being performed.

Although obtained with a simple model for the torus (\textit{i.e.},
with a constant specific angular momentum), the important feature that
distinguishes the hydrodynamic drag from radiation reaction, namely
the decrease of the inclination angle, cannot be affected by a change
of the specific angular momentum distribution (we recall that $\ell$
must be increasing with radius for stability). Such a feature, in
fact, is simply due to the conservation of the momentum of the
satellite in the $\theta$-direction during accretion and to the
dynamical friction of the fluid: both effects force the satellite to
smaller inclinations by reducing its $\theta$-velocity. However, the
calculation of the magnitude of the hydrodynamic drag and how it
compares with radiation reaction for more general disk models is not
straightforward. 

Tori built with increasing distributions of specific angular momentum,
in fact, would have two substantial differences with respect to those
considered here. Firstly, the separation between the specific angular
momentum of the torus and the Keplerian specific angular momentum will
generally decrease for orbits corotating with the torus, thus reducing
the relative motion between the satellite and the fluid and
consequently the hydrodynamic drag, whereas it will increase for
orbits counter-rotating relative to the torus, thus enhancing the
hydrodynamic drag. The magnitude of this effect depends on the precise
angular momentum distribution considered and rough estimates can be
made assuming a power-law for the specific angular momentum,
\textit{i.e.}, $\ell/M\sim (r/M)^{\alpha}$, with $\alpha<0.5$ for the
torus to have an outer radius and a
cusp~\cite{font_daigne,zanotti_etal:05}. Using the general formulas
reported in Sec.~\ref{sec:torus}, it is easy to check that for $r$
between $20 M$ and $50 M$ and prograde orbits the relative motion decreases by $\sim
20-30\%$ for $\alpha=0.1$ and by at least $95\%$ for $\alpha=0.4$
(this significant decrease is due to the fact that for $\alpha=0.4$
the center moves to a radius $r_{\rm center}\sim 27 M$, just in the
middle of the radial interval which we are considering). Conversely,
in the same radial range the increase for counter-rotating orbits is of
about $8-10\%$ ($30-45\%$), for $\alpha =0.1$ ($0.4$). Secondly, the
density in the inner parts of the torus will generally decrease. Using
again the expressions in Sec.~\ref{sec:torus}, it is easy to check
that the density decreases by about $18\%$ ($97\%$) at $r\sim20M$ and
by about $11\%$ ($90\%$) at $r\sim50M$, for $\alpha =0.1$ ($0.4$).

Overall, therefore, the decrease of the inclination angle due to the
hydrodynamic drag could be detectable by LISA also for non-constant
$\ell$ tori, especially if $\ell$ varies slowly with the radius or, if
$\ell$ varies rapidly with the radius, if the EMRI is counter-rotating
relative to the torus.

Finally, let us comment on two further effects that can in principle occur in the systems considered in
this paper. First, the motion of the satellite will be influenced by the gravitational attraction exterted by the torus. This
is clearly a conservative effect, and cannot therefore influence the infall of the satellite towards the SMBH, which is 
instead regulated by the dissipative forces (radiation reaction and hydrodynamic drag). However, this effect can in principle
cause the periastron to advance, thus introducing a phase-shift in the
gravitational waveforms. (Note that a similar advance is caused by the conservative part of the gravitational self-force~\cite{pound}.) To calculate the order of magnitude of this 
effect, let us consider for simplicity a thin disk of outer radius $r_{\rm out}$, mass 
$M_{\rm D}$ and constant surface density,
and a satellite of mass $m$ located on the equatorial plane
at a distance $d\ll r_{\rm out}$ from the central SMBH, the mass of which we denote by $M$. 
The potential energy of the satellite due to the gravitational field of the disk can be easily calculated to be, up to a constant and to leading order,
\eq 
U\approx\frac{m M_{\rm D}\, d^2}{2 r_{\rm out}^3}\,.
\eeq
This potential energy can be used to compute the Newtonian periastron precession of the 
satellite's orbit during a revolution [use for instance
eq.~(1) of Ref.~\cite{landau}, chapter 3, exercise number 3]:
\eq
\delta \phi\approx- \frac{3 M_{\rm D} \pi  d^3}{M r_{\rm out}^3}
\eeq 
for almost circular orbits. Using this equation and the well-known Netwonian formula for the revolution period, 
it is easy to check that, for orbits relevant for LISA, the total phase-shift accumulated 
in $1$ year is $\ll 2 \pi$ as long as $r_{\rm out} \gtrsim 10^4 M$ and $M=10^5-10^6 M_\odot$. 
Because LISA is not expected to detect phase lags less than 1 cycle over its lifetime\footnote{This corresponds indeed to a dephasing time of the order of LISA's lifetime.}, this periastron advance and the consequent phase-shift
cannot be observed.
On the other hand, for disks or tori with $r_{\rm out} \sim 10^3 M$, this effect could in
principle be marginally visible by LISA (especially if $M=10^5M_\odot$). 

A second effect which could in principle affect EMRIs 
in the presence of a torus is the spin of the satellite black hole, which increases due to accretion of the torus material.
The satellite's spin couples with the orbital angular momentum as well as 
with the spin of the SMBH, but its effect on the motion is negligible over a timescale of $1$ year~\cite{leor_spin},
unless it is close to its maximal value (in which case it might be marginally observable)~\cite{leor_spin}. 

\begin{acknowledgments}
It is a pleasure to thank L. Barack and J. C. Miller for giving helpful advice and
comments on this manuscript, P. Montero for insightful comments about
the torus solutions that we use in this paper, as well as L. Subr and
V. Karas for very useful remarks about the hydrodynamic interaction
between a moving body and a fluid. EB acknowledges the kind
hospitality of the Albert Einstein Institute where part of this work
was carried out.
\end{acknowledgments}

\end{document}